\documentclass{amsart}

\usepackage{amssymb}
\usepackage{graphicx}

\newtheorem{theorem}{Theorem}[section]

\newtheorem{proposition}[theorem]{Proposition}
\newtheorem{claim}[theorem]{Claim}

\newtheorem{example}[theorem]{Example}

\newtheorem{definition}[theorem]{Definition}

\numberwithin{equation}{section}
\newcommand{\Ai}{\text{Ai\,}}

\newcommand{\Tr}{\text{Tr\,}}
\newcommand{\sign}{\text{sgn\,}}
\newcommand{\im}{\text{Im\,}}

\begin{document}
\author{Kurt Johansson}
\address{Department of Mathematics, Royal Institute of technology,
SE-100 44 Stockholm, Sweden}

\email{kurtj@math.kth.se}

\thanks{Supported by the G\"oran Gustafsson Foundation (KVA)}

\title[Random matrices and determinantal processes]
{Random matrices and determinantal processes}

\maketitle    

\section{Introduction}\label{Introduction}

Eigenvalues of random matrices have a rich mathematical structure and are a source 
of interesting
distributions and processes. These distributions are natural statistical
models in many problems in quantum physics, \cite{Gu}. 
They occur for example, at least conjecturally, in
the statistics of spectra of quantized models whose classical dynamics is chaotic, 
\cite{BGS}.
Random matrix statistics is also seen 
in the statistics of zeros of L-functions in number theory, \cite{KaSa}.

In recent years we have seen a new development where probability distributions
from random matrix theory appear as limit laws in models
of a statistical mechanical nature, namely in certain 
random growth and random tiling problems.
This came as a surprise and has added a new side to random matrix theory.
It clearly shows that the limit probability measures coming out of
random matrix theory are natural limit probability distributions. 

Only very special models, which are exactly solvable in a certain sense, 
can be analyzed. In these notes we will survey two models, random domino tilings
of the Aztec diamond and a one-dimensional local random growth model, the corner
growth model. We will also discuss relations between these two models.
Underlying the exact solvability of the models is the fact that they
can be mapped to families of non-intersecting paths and that these in turn lead
to determinantal point processes. Point processes 
with determinantal correlation functions have emerged as an 
interesting class of point processes, with a rich structure and many interesting examples,
\cite{So}.

\section{Point processes}\label{Pointproc}
\subsection{General theory}

We will need some general facts about point processes, but we will only survey those
aspects that will be directly relevant for the present exposition, see \cite{DV-J}.
Let $\Lambda$ be a complete separable metric space and let $\mathcal{N}(\Lambda)$
denote the space of all counting measures $\xi$ on $\Lambda$ for which $\xi(B)<
\infty$ for every bounded $B\subseteq\Lambda$. We say that $\xi$ is {\it boundedly finite}. 
A counting measure is a measure $\xi$ whose values on bounded Borel sets in $\Lambda$ 
is a non-negative integer. Typically $\Lambda$ will be $\mathbb{R}$, $\mathbb{Z}$,
some subset of these or the disjoint union of
several copies of $\mathbb{R}$ or $\mathbb{Z}$. We can define a
$\sigma$-algebra $\mathcal{F}$ on $\mathcal{N}(\Lambda)$ by taking the smallest 
$\sigma$-algebra for which $A\to\xi(A)$ is measurable for all Borel sets $A$ in $\Lambda$.

If $B$ is a bounded set $\xi(B)$ is finite and we can write
\begin{equation}\label{1.1}
\left.\xi\right|_B =\sum_{i=1}^{\xi(B)}\delta_{x_i},
\end{equation}
for some $x_1,\dots,x_{\xi(B)}\in\Lambda$. Note that we can have $x_i=x_j$ for 
$i\neq j$, i.e. a multiple point. We say that $\xi$ is {\it simple} if $\xi(\{x\})
\le 1$ for all $x\in\Lambda$. The counting measure $\xi$ can be thought of as giving a 
point or particle configuration in $\Lambda$.
A {\it point process} on $\Lambda$ is a probability measure $\mathbb{P}$ on
$\mathcal{N}(\Lambda)$. The point process is {\it simple} if $\mathbb{P}(\xi
\,\,\text{simple})=1$.

If the function $\phi:\Lambda\to\mathbb{C}$ has support in the bounded Borel set
$B$ we write
\begin{equation}\label{1.2}
\prod_i (1+\phi(x_i))=\prod_{i=1}^{\xi(B)}(1+\phi(x_i)),
\end{equation}
where $x_1,\dots,x_{\xi(B)}$ are defined by (\ref{1.1}). If $\xi(B)=0$, then the
right hand side of (\ref{1.2}) is $=1$ by definition. Note that if $|\phi(x)|<1$
for all $x\in\Lambda$ we have
\begin{equation}\label{1.3}
\prod_i (1+\phi(x_i))=\exp(\int_{\Lambda}\log (1+\phi(x))d\xi(x)).
\end{equation}

A natural way to investigate a point process is to consider expectations of products of
the form (\ref{1.2}). If we take for instance $\phi=\exp(-\psi)-1$, $\psi\ge 0$ with 
bounded support, we get the so called Laplace functional. We can write
\begin{equation}\label{B1}
\prod_i (1+\phi(x_i))=\sum_{n=0}^\infty\frac 1{n!}\sum_{x_{i_1}\neq\dots\neq x_{i_n}}
\phi(x_{i_1})\dots\phi(x_{i_n}),
\end{equation}
where the sum is over all $n$-tuples of {\it distinct} points in the process and we
include all permutations of the $n$ points, which we compensate for by dividing by $n!$.
We want to include all permutations since there is no ordering of the points
in the process. If we have a multiple point of multiplicity $k$ it should be counted as
$k$ distinct points occupying the same position. The $n=0$ term in (\ref{B1}) is $=1$ by
definition. Since $\phi$ has bounded support, the second sum in (\ref{B1}) is actually finite 
almost surely. We can construct a new point process $\Xi_n$ in $\Lambda^n$
for each $n\ge 1$, by setting
\begin{equation}\label{B2}
\Xi_n=\sum_{x_{i_1}\neq\dots\neq x_{i_n}}\delta_{(x_{i_1},\dots,x_{i_n})},
\end{equation}
i.e. each $n$-tuple of points in the original point process $\xi$, including all 
permutations of the points, gives rise to a point in the new process.

We define a measure $M_n$ on $\Lambda^n$ by setting
\begin{equation}\label{B3}
M_n(A)=\mathbb{E}\,[\Xi_n(A)],
\end{equation}
for each bounded Borel set $A\subseteq\Lambda^n$, i.e. $M_n(A)$ is the expected number
of $n$-tuples of distinct points that fall in $A$. Here we assume that the process is
such that all the $M_n$, $n\ge 1$, are well-defined, $M_n(A)<\infty$ for bounded $A$.
The measure $M_n$ is an intensity measure for $n$-tuples of distinct points in the
original process. The formula (\ref{B1}) can now be written
\begin{equation}\label{B4}
\prod_{i}(1+\phi(x_i))=\sum_{n=0}^\infty\frac 1{n!}\int_{\Lambda^n}\prod_{j=1}^n
\phi(x_j)\Xi_n(d^nx).
\end{equation}
Assume that 
\begin{equation}\label{B5}
\sum_{n=0}^\infty\frac {||\phi||_\infty^n}{n!}M_n(B^n)<\infty,
\end{equation}
where the bounded set $B$ contains the support of $\phi$. Since,
\begin{equation}
\sum_{n=0}^\infty\frac 1{n!}\mathbb{E}\,\left[\left|\int_{\Lambda^n}\prod_{j=1}^n
\phi(x_j)\Xi_n(d^nx)\right|\right]\le
\sum_{n=0}^\infty\frac {||\phi||_\infty^n}{n!}M_n(B^n)
\notag
\end{equation}
it follows from Fubini's theorem that
\begin{equation}\label{B6}
\mathbb{E}\,[\prod_{i}(1+\phi(x_i))]=\sum_{n=0}^\infty\frac 1{n!}\int_{\Lambda^n}\prod_{j=1}^n
\phi(x_j)M_n(d^nx).
\end{equation}

Consider the case when $\phi$ is a simple function
\begin{equation}\label{B6'}
\phi(x)=\sum_{j=1}^m a_j\chi_{A_j}(x)
\end{equation}
with $A_1,\dots,A_m$ disjoint, measurable subsets of a bounded set $B$. 
Note that since the $A_j$'s are disjoint we have
$$
1+t\phi(x)=\prod_{j=1}^m(1+ta_j)^{\chi_{A_j}(x)},
$$
where $|t|\le 1$,
and hence
\begin{equation}\label{1.9}
\prod_i (1+t\phi(x_i))=\prod_{j=1}^m(1+ta_j)^{\xi(A_j)}.
\end{equation}
Set $1/n!$=0 if $n<0$. Then, by the binomial theorem,
\begin{align}\label{1.10}
\prod_{j=1}^m(1+ta_j)^{\xi(A_j)}&=\prod_{j=1}^m\sum_{n_j=0}^{\xi(A_j)}
\frac{\xi(A_j)!}{n_j!(\xi(A_j)-n_j)!}(ta_j)^{n_j}
\notag\\
&=\sum_{n_1,\dots,n_m=0}^\infty \prod_{j=1}^m\frac{(ta_j)^{n_j}}{n_j!}
\prod_{j=1}^m\frac{\xi(A_j)!}{(\xi(A_j)-n_j)!}
\notag\\
&=\sum_{n=0}^\infty\frac {t^n}{n!}\sum_{n_1+\dots+n_m=n}\binom{n}{n_1\dots n_m}
\prod_{j=1}^m a_j^{n_j}\prod_{j=1}^m\frac{\xi(A_j)!}{(\xi(A_j)-n_j)!}.
\end{align}
If $t,a_1,\dots,a_m$ are all positive, it follows from Fubini's theorem that
\begin{align}\label{B7}
&\mathbb{E}\,[\prod_i(1+t\phi(x_i))]
\notag\\
&=\sum_{n=0}^\infty\frac {t^n}{n!}\sum_{n_1+\dots+n_m=n}\binom{n}{n_1\dots n_m}
\prod_{j=1}^m a_j^{n_j}
\mathbb{E}\,\left[\prod_{j=1}^m\frac{\xi(A_j)!}{(\xi(A_j)-n_j)!}\right].
\end{align}
On the other hand, by (\ref{B6}),
\begin{align}\label{B8}
&\mathbb{E}\,[\prod_i(1+\phi(x_i))]
=\sum_{n=0}^\infty\frac {t^n}{n!}\int_{\Lambda^n}\prod_{k=1}^n\left(\sum_{j=1}^m
a_j\chi_{A_j}(x_k)\right)M_n(d^nx)
\notag\\
&=\sum_{n=0}^\infty\frac {t^n}{n!}\sum_{n_1+\dots+n_m=n}\binom{n}{n_1\dots n_m}
\prod_{j=1}^m a_j^{n_j}M_n({A_1^{n_1}\times\dots A_m^{n_m}})
\end{align}
Hence, for any bounded, disjoint Borel sets $A_1,\dots,A_m$ in $\Lambda$, and $n_i$, 
$1\le i\le m$, such that $1\le n_i\le n$ and $n_1+\dots+m_m=n$,
\begin{equation}\label{1.4}
M_n(A_1^{n_1}\times\dots\times A_m^{n_m})=
\mathbb{E}\,\left[\prod_{i=1}^m \frac{(\xi(A_i))!}{(\xi(A_i)-n_i)!}\right].
\end{equation}
This can be used as an alternative definition of the measure $M_n$. If $X$ is a random 
variable, $\mathbb{E}\,(X^k)$ is the $k$:\,th moment of
$X$, and $\mathbb{E}\,(X(X-1)\dots(X-k+1))$ is the $k$'th {\it factorial moment}
of $X$. For this reason $M_n$ is called the factorial moment measure since, by (\ref{2.4}),
can be defined using joint factorial moments.

In many cases there is a natural reference measure $\lambda$ on $\Lambda$ like
Lebesgue measure on $\mathbb{R}$ or the standard counting measure on
$\mathbb{Z}$. We can then ask if the factorial moment measure $M_n$ has a
density with respect to $\lambda^n$ on $\Lambda^n$.

\begin{definition}\label{Def1.2}
If $M_n$ is absolutely continuous with respect to $\lambda^n$ on $\Lambda^n$,
i.e.
\begin{equation}\label{1.5}
M_n(A_1,\dots,A_n)=\int_{A_1\times\dots\times A_n}\rho_n(x_1,\dots,x_n)
d\lambda(x_1)\dots d\lambda(x_n)
\end{equation}
for all Borel sets $A_i$ in $\Lambda$, we call $\rho_n(x_1,\dots,x_n)$ the 
$n$'th {\it correlation function} or {\it correlation density}. They are also called
product densities.
\end{definition}

We will be dealing with point processes for which all correlation functions exist.
In many cases if we are given the correlation functions $(\rho_n)_{n\ge 1}$ the
process is uniquely determined. As can be guessed from above, the uniqueness problem
is closely related to the classical moment problem.

In the case of a simple point process on $\mathbb{R}$ we can get some intuition for the
correlation functions as follows. Let $A_i=[y_i,y_i+\Delta y_i]$, $1\le i\le n$,
be disjoint intervals. If the $\Delta y_i$ are small we expect there to be either one
or no particle in each $A_i$. Hence, typically, the product  $\xi(A_1)\dots\xi(A_n)$
is $1$ if there is exactly one particle in each $A_i$ and $0$ otherwise.
From (\ref{1.5}) we then expect
\begin{equation}\label{1.6}
\rho_n(y_1,\dots,y_n)=\lim_{\Delta y_i\to 0}
\frac{\mathbb{P}[\text{one particle in each $[y_i,y_i+\Delta y_i]$, $1\le i\le n$}]}
     {\Delta y_1\dots \Delta y_n}.
\end{equation}
Note that $\rho_n(y_1,\dots,y_1)$ is not a probability density. The function
$\rho_1(y)$ is the
density of particles at $y$, but since we have many particles the event of finding a 
particle at $y_1$ and the event of finding a particle at $y_2$ are not disjoint even
if $y_1\neq y_2$. We should think of $\rho_n(y_1,\dots,y_1)$ as particle densities not 
probability densities. It follows from the argument above that if we have a simple
point process on $\mathbb{Z}$ (or some other countable or finite set), then
$\rho_n(y_1,\dots,y_n)$ is exactly the probability of finding particles at
$y_1,\dots,y_n$. 

The next proposition follows from (\ref{B5}) and (\ref{B6}). The condition (\ref{1.7}) below
implies (\ref{B5}) and we get (\ref{B6}), which is exactly (\ref{2.8}) by the definition of 
the correlation functions.

\begin{proposition}\label{Prop1.3}
Consider a point process all of whose correlation functions exist. Let $\phi$ be
a complex-valued, bounded, measurable function with bounded support. Assume that
the support of $\phi$ is contained in the bounded, measureable set $B$ and that
\begin{equation}\label{1.7}
\sum_{n=0}^\infty\frac{||\phi||_\infty^n}{n!}
\int_{B^n}\rho_n(x_1,\dots,x_n)d^n\lambda(x)<\infty.
\end{equation}
Then,
\begin{equation}\label{1.8}
\mathbb{E}\,[\prod_j (1+\phi(x_j))]=\sum_{n=0}^\infty\frac 1{n!}\int_{\Lambda^n}
\prod_{j=1}^n \phi(x_j)\rho_n(x_1,\dots,x_n)d^n\lambda(x).
\end{equation}
Here the product in the expectation in the left hand side is defined by (\ref{1.2}).
\end{proposition}

We can think of the left hand side of (\ref{1.8}) as a generating function for the 
correlation functions. Below we will see that (\ref{1.8}) is useful for computing
interesting probabilities. The condition (\ref{1.7}) is not intended to be optimal
but it will suffice for our purposes.

We also have a kind of converse of proposition \ref{Prop1.3}.
\begin{proposition}\label{Prop1.4}
Let $(u_n)_{n\ge 1}$ be a sequence of measurable functions $u_n:\Lambda^n\to\mathbb{R}$.
Assume that for any simple, measurable function $\phi$ with bounded support, our point
process satisfies
\begin{equation}\label{1.11}
\mathbb{E}\,[\prod_i(1+\phi(x_i))]=\sum_{n=0}^\infty\frac 1{n!}
\int_{\Lambda^n}\prod_{j=1}^n \phi(x_j)u_n(x_1,\dots,x_n)d\lambda^n(x),
\end{equation}
with a convergent right hand side. Then all the correlation functions $\rho_n$,
$n\ge 1$, exist and $\rho_n=u_n$.
\end{proposition}

\begin{proof}
Arguing as above in (\ref{B7}) and (\ref{B8}) we see that
\begin{align}
M_n(A_1^{n_1}\times\dots\times A_m^{n_m})&=\mathbb{E}\,\left[
\prod_{j=1}^m\frac{\xi(A_j)!}{(\xi(A_j)-n_j)!}\right]
\notag\\
&=\int_{A_1^{n_1}\times\dots\times A_m^{n_m}} u_N(x_1,\dots,x_n)d^n\lambda(x).
\end{align}
This proves the proposition by the definition of the correlation functions and
(\ref{1.4}).
\end{proof}

Proposition \ref{Prop1.3} is useful when we want to compute {\it gap probabilities}, i.e. 
the probability that there is no particle in a 
certain set. If $B$ is a bounded, measurable 
set and (\ref{1.7}) holds with $\phi=-\chi_B$, then
\begin{equation}\label{1.13}
\mathbb{P}[\text{no particle in $B$}]=\mathbb{E}\,[\prod_j(1-\chi_B(x_j))]=
\sum_{n=0}^\infty\frac{(-1)^n}{n!}\int_{B^n}\rho_n(x_1,\dots,x_n)d^n\lambda(x)
\end{equation}
Below we will be interested in processes on $\mathbb{R}$, or a subset of $\mathbb{R}$, 
which have a last or rightmost particle. Consider a point process $\xi$ on
$\mathbb{R}$. If there is a $t$ such that $\xi(t,\infty)<\infty$, we say that $\xi$
has a {\it last particle}. This will then be true for all $t$, since $\xi(A)<\infty$ for any
bounded set. If $x_1\le\dots\le x_{n(\xi)}$ are the finitely many particles in
$(t,\infty)$, we define
$x_{\max}(\xi)=x_{n(\xi)}$,
the {\it position of the last particle}. The distribution function $\mathbb{P}[x_{\max}(\xi)
\le t]$ is called the {\it last particle distribution}. If $\mathbb{E}\,[\xi(t,\infty)]
<\infty$ for some $t\in\mathbb{R}$, then $\xi$ has a last particle almost surely.

\begin{proposition}\label{Prop1.5}
Consider a point process $\xi$ on $\mathbb{R}$ or a subset of $\mathbb{R}$, all whose
correlation functions exist, and assume that
\begin{equation}\label{1.14}
\sum_{n=0}^\infty\frac{1}{n!}\int_{(t,\infty)^n}\rho_n(x_1,\dots,x_n)d^n\lambda(x)
<\infty
\end{equation}
for each $t\in\mathbb{R}$. Then the process $\xi$ has a last particle and
\begin{equation}\label{1.15}
\mathbb{P}[x_{\max}(\xi)\le t]=
\sum_{n=0}^\infty\frac{(-1)^n}{n!}\int_{(t,\infty)^n}\rho_n(x_1,\dots,x_n)d^n\lambda(x).
\end{equation}
\end{proposition}

\begin{proof} It follows from the $n=1$ term in (\ref{1.14}) that 
$\mathbb{E}\,[\xi(t,\infty)]<\infty$ and hence the process has a 
last particle almost surely. 
Take $t<s$. Proposition \ref{Prop1.3} implies that
\begin{equation}
\mathbb{P}[\text{no particle in $(t,s)$}]=
\sum_{n=0}^\infty\frac{(-1)^n}{n!}\int_{(t,s)^n}\rho_n(x_1,\dots,x_n)d^n\lambda(x).
\notag
\end{equation}
We see from (\ref{1.14}) and the dominated convergence theorem that we can let 
$s\to\infty$ and obtain (\ref{1.15}).
\end{proof}

Let us consider some examples of point processes.

\begin{example}\label{Ex1}
\rm A classical and basic example of a point process is the Poisson process on $\mathbb{R}$
with density $\rho(x)$, where $\rho$ is locally $L^1$. 
Let $A_1,\dots, A_m$ be disjoint, bounded sets in $\mathbb{R}$. Then $\xi(A_i)$ are
independent Poisson random variables with parameter $\int_{A_i}\rho(x)dx$, $1\le i\le m$.
Hence with $\phi$ as in (\ref{B6'}),
\begin{align}
&\mathbb{E}\,[\prod_{i}(1+\phi(x_i))]=
\prod_{j=1}^m\mathbb{E}\,[\prod_{j=1}^m(1+a_j)^{\xi(A_j)}]
\notag\\
&=\prod_{j=1}^m\left(\sum_{k=0}^\infty\frac{(1+a_j)^k}{k!}\left(
\int_{A_j}\rho(x)dx\right)^ke^{-\int_{A_j}\rho(x)dx}\right)
=e^{\int_{\mathbb{R}}\phi(x)\rho(x)dx}
\notag\\
&=\sum_{n=0}^\infty\frac 1{n!}\int_{\mathbb{R}^n}
\prod_{j=1}^n\phi(x_j)\prod_{j=1}^n\rho(x_j)d^nx.
\notag
\end{align}
It follows from proposition \ref{Prop1.4} that the correlation functions are given by
$$\rho_n(x_1,\dots,x_n)=\rho(x_1)\dots\rho(x_n),$$ which reflects the independence of 
particles at different locations. If $\rho(x)$ is integrable in $[t,\infty)$ the process
has a last particle almost surely and
\begin{equation}\label{1.16}
\mathbb{P}[x_{\max}(\xi)\le t]=
\sum_{n=0}^\infty\frac{(-1)^n}{n!}\int_{(t,\infty)^n}\rho(x_1)\dots\rho(x_n)d^n\lambda(x)=
\exp(-\int_t^\infty \rho(x)dx).\it
\end{equation}
\end{example}

\begin{example}\label{Ex2}
\rm
If $u_N(x_1,\dots,x_N)$ is a symmetric probability density on $\mathbb{R}^N$, then
$(x_1,\dots,x_N)\to\sum_{i=1}^N\delta_{x_i}$ maps the probability measure with density
$u_N$ to a finite point process on $\mathbb{R}$. The correlation functions are given
by
\begin{equation}\label{1.17}
\rho_n(x_1,\dots,x_n)=\frac{N!}{(N-n)!}\int_{\mathbb{R}^{N-n}}u_N(x_1,\dots,x_N)
dx_{n+1}\dots dx_N.
\end{equation}
i.e. they are multiples of the marginal densities. This is not difficult to see
using proposition \ref{Prop1.4}. When point processes defined in this way are studied 
(\ref{1.17}) is often taken as the definition of the correlation functions.
\it
\end{example}
\begin{example}\label{Ex3}
\rm
Let $\mathcal{H}_N$ be the space of all $N\times N$ Hermitian matrices. This space can
be identified with $\mathbb{R}^{N^2}$, since we have $N^2$ independent real numbers. If
$\mu_N$ is a probability measure on $\mathcal{H}_N$ and $\{\lambda_1(M),\dots,\lambda_N(M)
\}$ denotes the set of eigenvalues of $M\in\mathcal{H}_N$, then
\begin{equation}\label{1.18}
\mathcal{H}_N\ni M\to\sum_{j=1}^N\delta_{\lambda_j(M)}
\end{equation}
maps $\mu_N$ to a finite point process on $\mathbb{R}$.

If $dM$ is Lebesgue measure on $\mathcal{H}_N$, then
\begin{equation}\label{1.19}
d\mu_N(M)=\frac 1{\mathcal{Z}_N}e^{-\Tr M^2}dM
\end{equation}
is a Gaussian probability measure on $\mathcal{H}_N$ called the GUE ({\it
Gaussian Unitary Ensemble}). It can be shown, \cite{Me}, that for any symmetric,
continuous function on $\mathbb{R}^N$ with compact support
\begin{equation}
\int_{\mathcal{H}_N}f(\lambda_1(M),\dots,\lambda_N(M))d\mu_N(M)=
\int_{\mathbb{R}^N} f(x_1,\dots,x_N)u_N(x_1,\dots,x_N)d^Nx,
\end{equation}
where
\begin{equation}\label{1.20}
u_N(x_1,\dots,x_N)=\frac 1{Z_NN!}\prod_{1\le i<j\le N}(x_i-x_j)^2\prod_{j=1}^N
e^{-x_j^2}
\end{equation}
is the induced eigenvalue measure on $\mathbb{R}^N$. Hence the point process on $\mathbb{R}$
defined by (\ref{1.20})
has correlation functions given by (\ref{1.17}).\it
\end{example}

We will show below that the correlation functions for the GUE eigenvalue process have a 
particularly nice determinantal form. This leads us to introduce so-called determinantal
processes.

\subsection{Determinantal processes}

Determinantal processes are characterized by the fact that their correlation functions
have a certain determinantal form.

\begin{definition}
Consider a point process $\xi$ on a complete separable metric space $\Lambda$,
with reference measure $\lambda$, all of whose
correlation functions $\rho_n$ exist. If there is a function $K:\Lambda\times\Lambda\to
\mathbb{C}$ such that
\begin{equation}\label{1.21}
\rho_n(x_1,\dots,x_N)=\det (K(x_i,x_j))_{i,j=1}^n
\end{equation}
for all $x_1,\dots,x_n\in\Lambda$, $n\ge 1$, then we say that $\xi$ is a 
{\it determinantal point process}. We call $K$ the {\it correlation kernel} of the
process.  
\end{definition}

We can view the correlation kernel 
$K$ as an integral kernel of an operator $K$ on $L^2(\Lambda,\lambda)$,
\begin{equation}\label{1.22}
Kf(x)=\int_{\Lambda} K(x,y)f(y)d\lambda(y)
\end{equation}
provided the right hand side is well-defined.

Consider a determinantal process on $\Lambda$. Let $\phi\in L^\infty(\Lambda,\lambda)$
have bounded support in $B$. Then by proposition \ref{Prop1.3}

\begin{align}\label{1.23}
&\mathbb{E}\,[\prod_j(1+\phi(x_j))]=
\sum_{n=0}^\infty\frac 1{n!}\int_{\Lambda^n}\prod_{j=1}^n\phi(x_j)
\det(K(x_i,x_j))_{i,j=1}^n d^n\lambda(x)
\notag\\
&=\sum_{n=0}^\infty\frac 1{n!}\int_{B^n}\prod_{j=1}^n\phi(x_j)
\det(K(x_i,x_j))_{i,j=1}^n d^n\lambda(x)
\end{align}
provided
\begin{equation}\label{1.24}
\sum_{n=0}^\infty\frac {||\phi||_\infty^n}{n!}\int_{B^n}
\det(K(x_i,x_j))_{i,j=1}^n d^n\lambda(x)<\infty.
\end{equation}
The estimate (\ref{1.24}) can usually be proved using Hadamard's inequality.
The expansion (\ref{1.23}) can be taken as the definition of the Fredholm determinant
\linebreak
$\det(I+\chi_BK\chi_B\phi)_{L^2(\Lambda)}=\det(I+K\phi)_{L^2(B)}$. Here $K\phi$ is the
operator on $L^2(B)$ with kernel $K(x,y)\phi(y)$. There are other ways of defining the
Fredholm determinant for so called trace class operators, namely $\det(I+K\phi))_{L^2(B)}
=\prod_i(1+\lambda_i)$, where $\{\lambda_i\}$ are all the eigenvalues of the operator $K\phi$
on $L^2(B)$. If $K(x,y)\phi(y)$ defines a trace class operator on $L^2(B)$
and $\Tr K\phi=\int_B K(x,x)d\lambda(x)$, then theses two definitions agree.
See \cite{GGK} for more on Fredholm determinants.

\begin{proposition}\label{Prop1.6}
Consider a determinantal point process $\xi$ on a subset $\Lambda$ of $\mathbb{R}$
with a hermitian correlation kernel $K(x,y)$, i.e. 
$K(y,x)=\overline{K(x,y)}$. Assume that $K(x,y)$ defines a trace class operator $K$
on $L^2(t,\infty)$
for each $t\in \mathbb{R}$, and that
\begin{equation}\label{1.26}
\Tr K=\int_t^\infty K(x,x)d\lambda(x)<\infty.
\end{equation}
Then $\xi$ has a last particle almost surely and
\begin{equation}\label{1.27}
\mathbb{P}[x_{\max}(\xi)\le t]=\det(I-K)_{L^2(t,\infty)}.
\end{equation}
\end{proposition}

\begin{proof}
This follows from proposition \ref{Prop1.5} and the above discussion provided we can prove
(\ref{1.24}). Since correlation functions are non-negative and $K(y,x)=\overline{K(x,y)}$,
the matrix $(K(x_i,x_j))$ is positive definite. In that case Hadamard's inequality
says that $\det(K(x_i,x_j))_{1\le i,j\le n}\le\prod_{j=1}^nK(x_j,x_j)$. Hence
\begin{align}
\sum_{n=0}^\infty\frac {C^n}{n!}\int_{(t,\infty)^n}\det(K(x_i,x_j))_{1\le i,j\le n}d^n\lambda
(x)&\le\sum_{n=0}^\infty\frac {C^n}{n!}\left(\int_t^\infty K(x,x)d\lambda(x)\right)^n
\notag\\
&=\exp\left( C\int_t^\infty K(x,x)d\lambda(x)\right),
\notag
\end{align}
by (\ref{1.26}).
\end{proof}

Consider for example the {\it Airy kernel},
\begin{equation}\label{1.28}
A(x,y)=\int_0^\infty \Ai(x+t)\Ai (y+t)dt.
\end{equation}
This is a Hermitian kernel and $\int_t^\infty A(x,x)dx<\infty$ for any real $t$. 
It can be shown that $A(x,y)\chi_{(t,\infty)}$ is the kernel of a trace class operator, 
and that there is a point process $\xi$ on $\mathbb{R}$, the {\it Airy kernel point
process}, which is determinantal with
kernel $A(x,y)$. This follows from general theory, see \cite{So}. We have
\begin{equation}\label{1.29}
F_{\text{TW}}(t)\doteq\mathbb{P}[x_{\max}(\xi)\le t]=\det(I-A)_{L^2(t,\infty)}.
\end{equation}
The distribution function $F_{\text{TW}}(t)$ for the last particle in the Airy kernel
point process is a natural scaling limit of certain finite determinantal point processes.
We call it the {\it Tracy-Widom distribution}, \cite{TW1}.

We will now look at some general ways of getting finite determinantal point processes.
It is possible to get interesting infinite point processes by looking at 
scaling limits of these.
These limiting point processes are typically of a few standard types, e.g. the Airy
kernel point process is obtained when we scale around the last particle in some finite
point processes on $\mathbb{R}$. These infinite point processes are natural scaling limits
and it is an interesting problem to understand how universal they are. 
In section \ref{Asymptotics} 
we will see the Airy kernel point process arising as a scaling limit of a finite point
process associated with a random domino tiling of the so-called Aztec diamond. It also
occurs as the scaling limit of GUE around the largest eigenvalue, see example \ref{Ex3.1}.

The following determinantal identity, we will call the {\it generalized 
Cauchy-Binet identity}. If we take $\Lambda=\{1,\dots,M\}$, $\lambda$ as counting
measure on $\Lambda$, $\phi_i(k)=a_{ik}$ and $\psi_i(k)=b_{ik}$, $M\ge N$, we get
the classical Cauchy-Binet identity.

\begin{proposition}\label{Prop1.7}
Let $(\Lambda,\mathcal{B},\lambda)$ be a measure space, and let $\phi_j,\psi_j$, 
$1\le i,j\le N$, be measurable functions such that $\phi_i\psi_j$ is integrable for
any $i,j$. Then,
\begin{align}\label{1.30}
&\det\left(\int_\Lambda \phi_i(x)\psi_j(x)d\lambda(x)\right)_{1\le i,j\le N}
\notag\\
&=\frac 1{N!}\int_{\Lambda^N} \det(\phi_i(x_j))_{1\le i,j\le N}
\det(\psi_i(x_j))_{1\le i,j\le N} d^N\lambda(x).
\end{align}
\end{proposition}

\begin{proof}
This is a computation,
\begin{align}
&\det\left(\int_\Lambda \phi_i(x)\psi_j(x)d\lambda(x)\right)_{1\le i,j\le N}
=\int_{\Lambda^N} \det(\phi_i(x_i)\psi_j(x_i))_{1\le i,j\le N}d^N\lambda(x)
\notag\\
&=\int_{\Lambda^N}\prod_{i=1}^N\phi_i(x_i)\det(\psi_j(x_i))_{1\le i,j\le N}
d^N\lambda(x)
\notag\\
&=\int_{\Lambda^N}\prod_{i=1}^N\phi_i(x_{\sigma(i)})
\det(\psi_j(x_{\sigma(i)}))_{1\le i,j\le N}
d^N\lambda(x)
\notag\\
&=\int_{\Lambda^N}\sign(\sigma)\prod_{i=1}^N\phi_i(x_{\sigma(i)})
\det(\psi_j(x_i))_{1\le i,j\le N}
d^N\lambda(x)
\notag\\
&=\frac 1{N!}\int_{\Lambda^N}\sum_{\sigma\in S_N}\sign(\sigma)
\prod_{i=1}^N\phi_i(x_{\sigma(i)})
\det(\psi_j(x_i))_{1\le i,j\le N}
d^N\lambda(x).
\notag
\end{align}
The first equality follows immediately using the definition of the determinant. In the third
we have permuted the variables using an arbitrary permutation $\sigma\in S_N$, and in the
fourth equality we used the antisymmetry of the determinant. The last equality follows
since the integral is independent of $\sigma$. The final expression is exactly what we
want by the definition of the determinant.
\end{proof}

Consider now the measure
\begin{equation}\label{1.31}
u_N(x)d^N\lambda(x)=\frac 1{N!Z_N}
\det(\phi_i(x_j))_{1\le i,j\le N}\det(\psi_i(x_j))_{1\le i,j\le N} d^N\lambda(x)
\end{equation}
on $\Lambda^N$, where
\begin{equation}\label{1.32}
Z_N=\frac 1{N!}\int_{\Lambda^N} \det(\phi_i(x_j))_{1\le i,j\le N}
\det(\psi_i(x_j))_{1\le i,j\le N} d^N\lambda(x),
\end{equation}
and we assume that $Z_N\neq 0$. If $u_N(x)\ge 0$, then (\ref{1.31}) is a probability
measure on $\Lambda^N$. It follows from the generalized Cauchy-Binet identity
(\ref{1.30}) that $Z_N=\det A$, where
$A=(a_{ij})_{1\le i,j\le N}$, and
\begin{equation}\label{1.33}
a_{ij}=\int_\Lambda\phi_i(x)\psi_j(x)d\lambda(x).
\end{equation}

\begin{proposition}\label{Prop1.8}
Let $(\Lambda,\mathcal{B},\lambda)$ be a measure space and let $\phi_i,\psi_i$ be as
in proposition \ref{Prop1.7}. Assume that $Z_N$ given by (\ref{1.32})  is $\neq 0$.
Then the matrix $A$ defined by (\ref{1.33}) is invertible and we can define
\begin{equation}\label{1.34}
K_N(x,y)=\sum_{i,j=1}^N \psi_i(x)(A^{-1})_{ij}\phi_j(y).
\end{equation}
If $g\in L^\infty(X)$, we have the following identity
\begin{equation}\label{1.35}
\int_{\Lambda^N}\prod_{j=1}^N(1+g(x_j))u_N(x)d^N\lambda(x)=
\sum_{n=0}^N\frac 1{n!}\int_{\Lambda^n}\prod_{j=1}^n g(x_j)\det(K_N(x_i,x_j))_
{1\le i,j\le n} d^n\lambda(x),
\end{equation}
with $u_N(x)$ given by (\ref{1.31}).
\end{proposition}

\begin{proof}
That $A$ is invertible follows from the fact that $\det A=Z_N\neq 0$ by (\ref{1.30}). The proof 
of (\ref{1.35}) is based on the determinant expansion
\begin{equation}\label{1.36}
\det(I+C)=\sum_{n=0}^N\frac 1{n!}\sum_{i_1,\dots,i_n=1}^N\det(C_{i_ri_s})_{1\le r,s\le n},
\end{equation}
where $C$ is an arbitrary $N\times N$-matrix, and the formula (\ref{1.30}). The identity
(\ref{1.36}) is a consequence of multilinearity of the determinant and expansion along
columns. It follows from (\ref{1.30}) and (\ref{1.33}) that
\begin{align}\label{1.37}
&\int_{\Lambda^N}\prod_{j=1}^N(1+g(x_j))u_N(x)d^N\lambda(x)
\notag\\
&=\frac{\det(a_{jk}+\int_\Lambda \phi_j(x)\psi_k(x) g(x)d\lambda(x))}
{\det(a_{jk})}
\notag\\
&=\det\left(\delta_{jk}+\sum_{i=1}^N(A^{-1})_{ji}\int_\Lambda \phi_i(x)\psi_k(x)g(x)d\lambda(x)
\right)
\notag\\
&=\det\left(\delta_{jk}\int_\Lambda f_j(x)h_k(x)d\lambda(x)\right),
\end{align}
where $f_j(x)=\sum_{i=1}^N(A^{-1})_{ji}\phi_i(x)$ and $h_k(x)=g(x)\psi_k(x)$.
Using (\ref{1.36}) we see that the last expression in (\ref{1.37}) can be written
\begin{align}
&\sum_{n=0}^N\frac 1{n!}\sum_{i_1,\dots,i_n=1}^N\det\left(\int_\Lambda f_{i_j}(x)
h_{i_k}(x)d\lambda(x)\right)_{1\le j,k\le n}
\notag\\
&=\sum_{n=0}^N\frac 1{n!}\sum_{i_1,\dots,i_n=1}^N\frac 1{n!}\int_{\Lambda^n}
\det(f_{i_j}(x_k))_{1\le j,k\le n}\det(h_{i_j}(x_k))_{1\le j,k\le n}d^n\lambda(x)
\notag\\
&=\sum_{n=0}^N\frac 1{n!}
\int_{\Lambda^n}\det\left(\sum_{i=1}^Nf_i(x_j)h_i(x_k))d^n\lambda(x)\right)_{1\le j,k\le n},
\notag
\end{align}
where we have used the identity (\ref{1.30}) in the two last equalities. Since
\begin{equation}
\sum_{j=1}^Nf_j(x)h_j(x)=g(x)\sum_{i,j=1}^N\psi_j(x)(A^{-1})_{ji}\phi_i(y)
\notag
\end{equation}
we are done.
\end{proof}

Assume now that $\Lambda$ is a complete separable metric space. If $u_N(x)\ge 0$, then 
(\ref{1.31}) is a probability measure on $\Lambda^N$ and the map $\Lambda^N\ni (x_1,\dots,x_N)
\to\sum_{j=1}^N\delta_{x_j}$ maps this to a point process $\xi$ on $\Lambda$. It follows
from proposition \ref{Prop1.4} and the identity (\ref{1.35}) that $\xi$ is a determinantal
point process with correlation functions given by (\ref{1.34}). Although (\ref{1.34}) gives
an explicit formula for the correlation kernel it is rather complicated. In particular, if we 
want to study a scaling limit as $N\to\infty$, we have to be able to find the inverse
of the $N\times N$-matrix $A$ in a useful form. Sometimes it is possible to do row
operations in the two determinants in (\ref{1.31}) so that the matrix $A$ becomes
diagonal and hence trivial to invert.

\begin{example}\label{Ex3.1} (The orthogonal polynomial method). \rm 
Consider the GUE eigenvalue
measure (\ref{1.20}). The density can be written as
\begin{equation}\label{1.38}
u_N(x_1,\dots,x_N)=\frac 1{Z_NN!}\det(x_i^{j-1}e^{-x_i^2/2})_{1\le i,j\le N}^2.
\end{equation}
If $p_j(x)$ is an arbitrary polynomial of degree $j$, $j=0,1,\dots$, then by doing row
operations in the determinant we see that
\begin{equation}\label{1.39}
u_N(x_1,\dots,x_N)=\frac 1{Z_N'N!}\det(p_{j-1}(x_i)e^{-x_i^2/2})_{1\le i,j\le N}^2.
\end{equation}
It now follows from proposition \ref{Prop1.8} 
that the GUE eigenvalue process has determinantal
correlation functions. The elements in the matrix $A$ are given by
\begin{equation}
a_{ij}=\int_{\mathbb{R}} p_{i-1}(x)p_{j-1}(x)e^{-x^2}dx.
\end{equation}
It is clear that it is very natural to choose $p_j$ to be the $j$\,th normalized Hermite 
polynomial so that $a_{ij}=\delta_{ij}$. The correlation kernel is then given by
\begin{equation}\label{1.40}
K_N(x,y)=\sum_{j=0}^{N-1}p_j(x)p_j(y)e^{-(x^2+y^2)/2}.
\end{equation}

We obtain the Airy kernel point process in the large $N$ limit when we scale around
the largest eigenvalue of an $N\times N$ random matrix from the GUE. More precisely,
let $\lambda^{(N)}_1\ge\lambda^{(N)}_2\ge\dots\ge\lambda^{(N)}_N$ be the eigenvalues
and set
$$
x_j=\frac{\sqrt{2N}\lambda^{(N)}_j-2N}{N^{1/3}},
$$
$j\ge 1$. Under the GUE eigenvalue measure $\xi_N=\sum_{j=1}^N\delta_{x_j}$
becomes a point process and in the limit $N\to\infty$ this process converges to the
Airy kernel point process. The proof is based on the fact that we can investigate
the scaling limit of the correlation kernel (\ref{1.40}) using asymptotics of
Hermite polynomials. Also, if $\lambda_{\max}(N)$ ($=\lambda^{(N)}_1$) denotes
the largest eigenvalue then proposition \ref{Prop1.6} can be used to show that
$$
\lim_{N\to\infty}\mathbb{P}\left[\frac{\sqrt{2N}\lambda_{\max}(N)-2N}{N^{1/3}}\le t
\right]=F_{\text{TW}}(t).
$$
From this example, which has several generalizations, we see that orthogonal polynomial
asymptotics is important in studying the asymptotics of the eigenvalues in some random
matrix ensembles.
\it
\end{example}

\subsection{Measures defined by products of several determinants}
There is a useful extension of proposition \ref{Prop1.8} to the case when the measure is 
given by a product of several determinants. Later we will see that such measures 
arise naturally in interesting problems. 
Actually both our main models will be of this type.
Let $X$ be a complete separable metric space with
a Borel measure $\mu$ and fix $m,n\ge 1$. Furthermore, 
let $\phi_{r,r+1}:X\times X\to\mathbb{C}$, $r=1,\dots,m-1$ be given measurable {\it transition
functions}, and $\phi_{0,1}:X_0\times X\to\mathbb{C}$, 
$\phi_{m,m+1}:X\times X_{m+1}\to\mathbb{C}$ given {\it initial} and {\it final}
transition functions. Here $X_0$ and $X_{m=1}$ are some given sets, which could be $X$
or $\{1,\dots,n\}$ for example. We will consider measures on $(X^n)^m$ of the form
\begin{align}\label{1.41}
&p_{n,m}(\underline{x})d\mu(\underline{x})=\frac 1{(n!)^mZ_{n,m}}w_{n,m}(\underline{x})
d\mu(\underline{x})
\notag\\
&\doteq\frac 1{(n!)^mZ_{n,m}}\det(\phi_{0,1}(x^0_i,x^1_j))
\prod_{r=1}^{m-1}\det(\phi_{r,r+1}(x^r_i,x^{r+1}_j))
\det(\phi_{m,m+1}(x^m_i,x^{m+1}_j))d\mu(\underline{x}),
\end{align}
where $\underline{x}=(x^1,\dots,x^m)\in(X^n)^m$, $x^r=(x_1,\dots,x_n^r)$,
$d\mu(\underline{x})=\prod_{r=1}^m\prod_{j=1}^nd\mu(x_j^r)$ and $x^0\in X_0^n$,
$x^{m+1}\in X_{m+1}^n$ are fixed points.
Here,
\begin{equation}\label{1.42}
Z_{n,m}=\frac 1{(n!)^m}\int_{(X^n)^m}w_{n,m}(\underline{x})d\mu(\underline{x})
\end{equation}
and we assume that $Z_{n,m}\neq 0$.
If $p_{n,m}(\underline{x})\ge $ we get a probability measure on $(X^n)^m$. Set
$\Lambda=\{1,\dots,m\}\times X$. If we map $x^r_j$ to $(r,x^r_j)\in\Lambda$, 
$\underline{x}$ gives $N=mn$ points in $\Lambda$ with exactly $n$ points in $\{r\}
\times X$ for each $r$. In this way we get a point process $\xi$ on $\Lambda$ from 
the probability measure (\ref{1.41}). Let $\nu$ denote counting measure on $\{1,\dots,m\}$. 
We will use $\lambda=\nu\times\mu$ as our reference measure on $\Lambda$.
When $m=1$ we can take $X_0=X_{m+1}=\{1,\dots, n\}$, $\phi_{0,1}(i,x)=\phi_i(x)$
and $\phi_{m,m+1}(x,j)=\psi_j(x)$ to obtain the measure (\ref{1.31}), so the present setting 
generalizes the one considered above. Our aim is to show that this more general setting
also leads to a determinantal process. Variants of this type of setting have been developed in
\cite{EyMe}, \cite{FS}, \cite{FNH} and \cite{JDPNG}.

Given two transition functions $\phi,\psi$ we define their composition by
$\phi\ast\psi(x,y)=\int_{X}\phi(x,z)\psi(z,y)d\mu(z)$. Set
\begin{equation}
\phi_{r,s}(x,y)=(\phi_{r,r+1}\ast\dots\ast\phi_{s-1,s})(x,y),
\notag
\end{equation}
when $r<s$ and $\phi_{r,s}\equiv 0$ if $r\ge s$. We assume that the transition functions are
such that all functions $\phi_{r,s}$, $0\le r<s\le m+1$ are well-defined. This will imply
that the integral in (\ref{1.42}) is convergent as can be seen by expanding the determinants.
Set $A=(a_{ij})$, where
\begin{equation}\label{1.43}
a_{i,j}=\phi_{0,m+1}(x^0_i,x^{m+1}_j).
\end{equation}
Repeated use of the generalized Cauchy-Binet identity (\ref{1.30}) gives
$Z_{n,m}=\det A$. Since we assume that $Z_{n,m}\neq 0$,
we see that $A$ is invertible. Set
\begin{equation}\label{1.44}
K_{n,m}(r,x;s,y)=\tilde{K}_{n,m}(r,x;s,y)-\phi_{r,s}(x,y),
\end{equation}
where $r,s\in\{1,\dots,m\}$, $x,y\in X$ and
\begin{equation}\label{1.45}
\tilde{K}_{n,m}(r,x;s,y)=\sum_{i,j=1}^n\phi_{r,m+1}(x,x^{m+1}_i)(A^{-1})_{ij}
\phi_{0,s}(x^0_j,y).
\end{equation}
We can now formulate the main result for measures of the form (\ref{1.41}).

\begin{proposition}\label{Prop1.9}
We use the notation above. Let $g:\Lambda\to\mathbb{C}$ belong to $L^\infty(\Lambda,\lambda)$
with support in a Borel set $B\subseteq\Lambda$. Let $\psi(r,x;s,y)=\chi_B(x)\phi_{r,s}
(x,y)g(s,y)$, $0\le r,s\le m+1$, where we omit $\chi_B(x)$ if $r=0$ and $g(s,y)$ if
$s=m+1$. Assume that $\psi$ defines a trace class operator, also denoted by $\psi$, on
$L^2(\Lambda,\lambda)$ which satisfies $\Tr \psi=\int_\Lambda\psi(z;z)\lambda(z)$. Then
\begin{align}\label{1.46}
&\int_{(X^n)^m}\prod_{r=1}^m\prod_{j=1}^n(1+g(r,x^r_j))p_{n,m}(\underline{x})d\mu
(\underline{x})=\det(I+\chi_BK_{n,m} g)_{L^2(\Lambda,\lambda)}
\notag\\
&=\sum_{k=0}^\infty\frac 1{k!}\int_{\Lambda^k}\prod_{j=1}^k g(z_j)
\det(K_{n,m}(z_i;z_j))_{1\le i,j\le k}d^k\lambda(z).
\end{align}
\end{proposition}

It follows from proposition \ref{Prop1.4} that the point process $\xi$ on $\Lambda$ is
determinantal with correlation kernel $K_{n,m}(z_1;z_2)$, $z_1,z_2\in\Lambda$.

\begin{proof} The kernel $\tilde{K}_{n,m}$ given by (\ref{1.45}) has finite rank and by
assumption $\psi$ is trace class. Hence $\chi_BK_{n,m}g$ is trace class and the Fredholm
determinant $\det(I+\chi_BK_{n,m}g)$ is well defined. Since also
$\Tr(\chi_BK_{n,m}g)=\int_\Lambda(\chi_BK_{n,m}g)(z;z)d\lambda(z)$, this Fredholm determinant
has an expansion as given in the theorem, \cite{GGK}. Write
\begin{equation}
Z_{n,m}[g]=\frac 1{(n!)^m}\int_{(X^n)^m}
\prod_{r=1}^m\prod_{j=1}^n(1+g(r,x^r_j))w_{n,m}(\underline{x})d\mu(\underline{x}),
\notag
\end{equation}
so that $Z_{n,m}[0]=Z_{n,m}=\det A$. Repeated use of the generalized Cauchy-Binet identity
(\ref{1.30}) gives
\begin{align}
&Z_{n,m}[g]
\notag\\
&=\det\left(\int_{X^m}\phi_{0,1}(x^0_i,t_1)\prod_{r=1}^m (1+g(r,t_r))
\prod_{r=1}^{m-1}\phi_{r,r+1}(t_r,t_{r+1})\phi_{m,m+1}(t_m,x^{m+1}_j)
d^m\mu(t)\right)_{1\le i,j\le n}.
\notag
\end{align}
We can write
\begin{equation}
\prod_{r=1}^m (1+g(r,t_r))=1+\sum_{\ell=1}^m\sum_{1\le r_1<\dots<r_\ell\le m}
g(r_1,t_{r_1})\dots g(r_\ell,t_{r_\ell})
\notag
\end{equation}
and thus
\begin{align}\label{1.47}
Z_{n,m}[g]&=\det\left(a_{ij}+\sum_{\ell=1}^m\sum_{1\le r_1<\dots<r_\ell\le m}
\int_{X^\ell}d^\ell\mu(t)\phi_{0,r_1}(x^0_i,t_1)g(r_1,t_1)\right.
\notag\\
&\times\prod_{s=1}^{\ell-1}\phi_{r_s,r_{s+1}}(t_s,t_{s+1})
g(r_{s+1},t_{s+1})\phi_{r_\ell,m+1}(t_\ell,x^{m+1}_j)\left.\right).
\end{align}
By definition $\phi_{r,s}=0$ if $r\ge s$ and hence we can remove the ordering of 
the $r_i$'s in (\ref{1.47}). We obtain
\begin{align}\label{1.48}
\frac{Z_{n,m}[g]}{Z_{n,m}[0]}=&\det\left(\delta_{i,j}+\sum_{k=1}^n(A^{-1})_{ik}
\sum_{\ell=1}^m\sum_{r_1,\dots,r_\ell}^m\int_{X^\ell} 
d^\ell\mu(t)\phi_{0,r_1}(x^0_k,t_1)g(r_1,t_1)\right.
\notag\\
&\times\left.\prod_{s=1}^{\ell-1}
\psi(r_s,t_s;r_{s+1},t_{s+1})\phi_{r_\ell,m+1}(t_\ell,j)\right)
\notag\\
&=\det\left(\delta_{i,j}+\sum_{k=1}^n(A^{-1})_{ik}\int_\Lambda d\lambda(r,x)
\int_\Lambda d\lambda(s,y)\phi_{0,r}(x^0_k,x)g(r,x)\right.
\notag\\
&\times\left.\left(\sum_{\ell}^m\psi^{\ast(\ell-1)}(r,x;s,y)\right)
\phi_{s,m+1}(y,x^{m+1}_j)\right),
\end{align}
where $\psi^{\ast 0}(r,x;s,y)=\delta_{r,s}\delta(x-y)$ and recursively
$$
\psi^{\ast\ell}(r,x;s,y)=\int_\Lambda \psi(r,x;u,t)\psi^{\ast(\ell-1)}(u,t;s,y)
d\lambda(u,t)
$$
for $\ell\ge 1$.

Set
\begin{align}
b(i;r,x)&=\sum_{k=1}^n(A^{-1})_{ik}\phi_{0,r}(x^0)_k,x)g(r,x)
\notag\\
c(r,x;j)&=\int_\Lambda\left(\sum_{\ell}^m\psi^{\ast(\ell-1)}(r,x;s,y)\right)\psi_{s,m+1}(y,x^{m+1}_j)
d\lambda(s,y)
\notag
\end{align}
and let $b:L^2(\Lambda,\lambda)\to\ell^2(n)$, $c:\ell^2(n)\to L^2(\Lambda,\lambda)$ denote
the corresponding operators. Then, by (\ref{1.48}),
\begin{equation}
\frac{Z_{n,m}[g]}{Z_{n,m}[0]}=\det(\delta_{ij}+(bc)(i,j))_{1\le i,j\le n}
=\det(I+cb)_{L^2(\Lambda,\lambda)}.
\notag
\end{equation}
Now,
\begin{equation}
cb=\left(\sum_{\ell=1}^m\psi^{\ast(\ell-1)}\right)(\chi_B\tilde{K}g).
\notag
\end{equation}
(The insertion of the $\chi_B$ does not change anything.)  By assumption $\psi$
is a trace class operator and using $\phi_{r,s}\equiv 0$ if $r\ge s$, we see that it is 
nilpotent, $\psi^{\ast\ell}\equiv 0$ if $\ell\ge m$. Hence $\det(I-\psi)=1$.
Consequently,
\begin{align}
\frac{Z_{n,m}[g]}{Z_{n,m}[0]}&=\det(I-\psi)\det(I+
(\sum_{\ell=0}^{m-1}\psi^{\ast(\ell)})(\chi_B\tilde{K}g))
\notag\\
&=\det(I-\psi+\chi_B\tilde{K}g)=\det(I+\chi_BKg),
\notag
\end{align}
and we are done.
\end{proof}

\section{Non-intersecting paths and the Aztec diamond}\label{SectAztec}

\subsection{Non-intersecting paths and the LGV theorem}
A natural way to obtain measures of the form (\ref{1.31}) and (\ref{1.41}) is from
non-intersecting paths. This is a consequence of the Lindstr\"om-Gessel-Viennot theorem
in the discrete setting, \cite{Stem}, and the Karlin-McGregor theorem in the case of
non-colliding continuous Markov processes in one-dimension. In our applications below we
will use the discrete setting so we will concentrate on that.

Let $\mathcal{G}=(V,E)$ be a directed acyclic graph with no multiple edges.
A {\it directed path} $\pi$ from a vertex $u$ to a vertex $v$ in $\mathcal {G}$
is a sequence of vertices $x_1,\dots, x_m$ in $\mathcal{G}$ such that $x_ix_{i+1}$, 
the edges in the path, are directed edges in $\mathcal{G}$, $x_1=u$ and $x_m=v$. The set 
of all directed paths from $u$ to $v$ will be denoted by $\Pi(u,v)$. If $u_1,\dots, u_n$ and
$v_1,\dots,v_n$ are vertices in $\mathcal{G}$, then $\Pi(\mathbf{u},\mathbf{v})$,
$\mathbf{u}=(u_1,\dots,u_n)$, $\mathbf{v}=(v_1,\dots,v_n)$, denotes the set of all directed 
paths $(\pi_1,\dots,\pi_n)$, where $\pi_i$ is a directed path from $u_i$ to $v_i$, 
$1\le i\le n$.
We say that two directed paths {\it intersect} if they share a common vertex. The families of 
paths in $\Pi(\mathbf{u},\mathbf{v})$ that do not have any intersections with each other
is denoted by $\Pi_{\text{n.i.}}(\mathbf{u},\mathbf{v})$, and those that have at least
one intersection by $\Pi_{\text{w.i.}}(\mathbf{u},\mathbf{v})$. If $\sigma\in S_n$ is a
permutation of $\{1,\dots,n\}$ we will write $\mathbf{v}_{\sigma}$ for
$(v_{\sigma(1)},v_{\sigma(2)},\dots,v_{\sigma(n)})$.

Let $w:E\to\mathbb{C}$ be a given function, called the {\it weight function}, $w(e)$ is the
{\it weight} of the edge $e$ in $\mathcal{G}$. The weight of a path is $w(\pi)=
\prod_{e\in\pi} w(e)$, i.e. the product of the weights over all edges in the path. The
weight of paths $(\pi_1,\dots,\pi_n)$ from $\mathbf{u}$ to $\mathbf{v}$ is
$w(\pi_1,\dots,\pi_n)=w(\pi_1)\dots w(\pi_n)$.
If $S\subseteq\Pi(\mathbf{u},\mathbf{v})$, then the weight of the set $S$ is
\begin{equation}\label{1.49}
W(S)=\sum_{(\pi_1,\dots,\pi_n)\in S}w(\pi_1,\dots,\pi_n).
\end{equation}
The total weight of all paths between two vertices $u$ and $v$ will be denoted by
\begin{equation}\label{1.50}
\phi(u,v)=W(\Pi(u,v))=\sum_{p\in\Pi(u,v)} w(\pi).
\end{equation}
We will call $\phi(u,v)$ the {\it transition weight} from $u$ to $v$.
Here we are assuming that the sum in the right hand side of (\ref{1.50}) is convergent. 
We could also
regard the weights as formal variables in some ring and (\ref{1.51}) as an identity in that ring.

We can now formulate the Lindstr\"om-Gessel-Viennot (LGV) theorem which relates 
weights of non-intersecting paths and determinants.

\begin{theorem} Let $\mathcal{G}$ be a directed, acyclic graph and $\mathbf{v}=
(u_1,\dots,u_n)$, $\mathbf{v}=(v_1,\dots,v_n)$ two $n$-tuples of vertices in $\mathcal{G}$ such 
that $\Pi_{\text{n.i.}}(\mathbf{u},\mathbf{v}_\sigma)\neq\emptyset$ only if $\sigma=\text{id}$.
Then,
\begin{equation}\label{1.51}
W(\Pi_{\text{n.i.}}(\mathbf{u},\mathbf{v}))=\det(\phi(u_i,v_j))_{i,j=1}^n.
\end{equation}
\end{theorem}
\begin{proof} (\cite{Stem}).
Expand the determinant in the right hand side of (\ref{1.51}). 
By (\ref{1.49}) and (\ref{1.50}),
\begin{align}
&\det(\phi(u_i,v_j))=\sum_{\sigma\in S_n} \text{sgn\,}(\sigma)\phi(u_1,v_{\sigma(1)})
\dots\phi(u_n,v_{\sigma(n)})
\notag\\
&=\sum_{\sigma\in S_n} \text{sgn\,}(\sigma)\prod_{i=1}^n W(\Pi(u_i,v_{\sigma(i)}))
=\sum_{\sigma\in S_n} \text{sgn\,}(\sigma)\prod_{i=1}^n 
\left(\sum_{\pi_i\in\Pi(u_i,v_{\sigma(i)})}w(\pi_i)\right)
\notag\\
&=\sum_{\sigma\in S_n} \text{sgn\,}(\sigma)\sum_{\pi\in\Pi(\mathbf{u},\mathbf{v}_\sigma)}
w(\pi_1,\dots,\pi_n)
\notag\\
&=\sum_{\sigma\in S_n} 
\sum_{\pi\in\Pi
_{\text{n.i.}}(\mathbf{u},\mathbf{v}_\sigma)}
\text{sgn\,}(\sigma)w(\pi_1,\dots,\pi_n)+
\sum_{\sigma\in S_n} \sum_{\pi\in\Pi
_{\text{w.i.}}(\mathbf{u},\mathbf{v}_\sigma)}
\text{sgn\,}(\sigma)w(\pi_1,\dots,\pi_n)
\notag\\
&\doteq S_1+S_2.
\notag
\end{align}
By assumption $\Pi_{\text{n.i.}}(\mathbf{u},\mathbf{v}_\sigma)=\emptyset$ unless $\sigma=
\text{id}$, and hence $S_1=W(\Pi_{\text{n.i.}}(\mathbf{u},\mathbf{v}))$. It remains to show
that $S_2=0$. 

Choose a fixed total order of the vertices, and let $\omega$ denote the first vertex in
this order which is a point of intersection between the paths $\pi_1,\dots,\pi_n$. Let
$\pi_i$ and $\pi_j$ be the two paths with smallest indices which intersect $\omega$.
Define a map
\begin{equation}\label{1.52}
(\sigma,\pi_1,\dots,\pi_n)\to(\sigma',\pi_1',\dots,\pi_n')
\end{equation}
as follows. Set $\pi_k'=\pi_k$ for $k\neq i,j$, and if
\begin{align}
\pi_i=u_ix_1\dots x_\alpha\omega x_{\alpha+1}\dots x_\beta v_{\sigma(i)}
\notag\\
\pi_j=u_jy_1\dots _\gamma\omega y_{\gamma+1}\dots y_\delta v_{\sigma(j)}
\notag
\end{align}
then
\begin{align}
\pi_i=u_ix_1\dots x_\alpha\omega y_{\gamma+1}\dots y_\delta v_{\sigma(j)}
\notag\\
\pi_j=u_jy_1\dots _\gamma\omega x_{\alpha+1}\dots x_\beta v_{\sigma(i)}.
\notag
\end{align}
Also, we set $\sigma'=\sigma\circ (i,j)$, where $(i,j)$ denotes the transposition of $i$
and $j$. Clearly, $w(\pi_1,\dots,\pi_n)=w(\pi_1',\dots,\pi_n')$ and $\text{sgn\,}(\sigma)
=-\text{sgn\,}(\sigma')$.

If we can show that (\ref{1.52}) is an involution, then $S_2=0$ follows since
\begin{align}
S_2&=\sum_{\sigma'\in S_n} \sum_{\pi'\in\Pi
_{\text{w.i.}}(\mathbf{u},\mathbf{v}_\sigma)}
\text{sgn\,}(\sigma')w(\pi_1',\dots,\pi_n')
\notag\\
&=-\sum_{\sigma\in S_n} \sum_{\pi\in\Pi
_{\text{w.i.}}(\mathbf{u},\mathbf{v}_\sigma)}
\text{sgn\,}(\sigma)w(\pi_1,\dots,\pi_n)=-S_2.
\notag
\end{align}
That (\ref{1.52}) is an involution is clear if $(\pi_1',\dots,\pi_n')$ has the same first
intersection point as $(\pi_1,\dots,\pi_n)$. Since only $\pi_i$ and $\pi_j$ were changed a
new intersection point has to occur between them. Assume that $x_m$ is the new intersection
point which is smallest in the total ordering. We must then have $x_m=x_\ell$, where one of 
$m$ and $\ell$ lies in $\{1,\dots,\alpha\}$ and the other in $\{\alpha+1,\dots,\beta\}$, say
$m$ lies in the first set. But then $x_m\dots x_\ell$ in $\pi_i$ is a cycle, which is 
impossible since we assumed that $\mathcal{G}$ is acyclic. Hence (\ref{1.52}) defines
an involution.
\end{proof}

We will see below that a combination of proposition \ref{Prop1.8}
or proposition \ref{Prop1.9} with the LGV-theorem will lead us to interesting
determinantal point processes in certain models.

\subsection{The Aztec diamond}

In this section we will discuss random domino
tilings of a region called the Aztec diamond. The model can equivalently be thought of
as a dimer model on a certain graph, \cite{Ke}. A typical tiling of the Aztec diamond
has the interesting feature that parts of it are completely regular, whereas the central part 
looks
more or less random. In fact there is a well defined random curve which separates the regular 
regions from the irregular region, and it is this curve that will be our main interest.
It turns out that the tiling can be described using non-intersecting paths in a certain
directed graph, and these paths will lead to a description of a random tiling by a
determinantal point process using the results of the previous section.

The {\it Aztec diamond}, $A_n$, of size $n$ is the union of all lattice squares
$[m,m+1]\times [\ell.\ell+1]$, $m,\ell\in\mathbb{Z}$, that lie inside the region
$\{(x_1,y_1)\,;\, |x_1|+|y_1|\le n+1\}$. A {\it domino} is a closed $1\times 2$ 
or $2\times 1$  rectangle in $\mathbb{R}^2$ with corners in $\mathbb{Z}^2$, and a
{\it tiling} of $A_n$ by dominos is a set of dominos whose interiors are disjoint
and whose union is $A_n$. Let $\mathcal{T}(A_n)$ denote the set of all domino tilings
of he Aztec diamond. The basic coordinate system used here will be referred to as
coordinate system I (CS-I).

We can color the unit squares in the Aztec diamond in a checkerboard fashion so that the
leftmost square in each row in the top half is white. Depending on how a domino covers
the colored squares we can distinguish four types of dominos. A horizontal domino is an 
{\it N-domino} if its leftmost square is white, otherwise it is an {\it S-domino}.
Similarly, a vertical domino is a {\it W-domino} if its upper square is white, otherwise it is
and {\it E-domino}. Two dominos are {\it adjacent} if they share an edge of a square, and a 
domino is {\it adjacent to the boundary} if it shares an edge with $\partial A_n$. We can
now define four regions where the tiling has a regular brick wall pattern. 
The {\it north polar region} (NPR) is defined to be the union of those N-dominos that are
connected with the boundary by a sequence of adjacent N-dominos, the last one of which is 
adjacent to the boundary.

Let $T\in\mathcal{T}(A_n)$ be a tiling of the Aztec diamond and let $v(T)$ denote the
number of vertical dominos in $T$. We define the weight of $T$ by letting vertical 
dominos have
weight $a$ and horizontal dominos weight 1, so that the total weight is $a^{v(T)}$. If 
$a>0$, which we assume, we get a probability measure on $\mathcal{T}(A_n)$ by normalizing this
weight. When $a=1$ we pick the tiling uniformly at random.

A tiling of the Aztec diamond with dominos can be described by a family of non-intersecting
paths. These paths can be obtained by drawing paths on the different types of dominos.
On an N-domino we draw no path. On a W-domino placed so that it has corners at
$(0,0)$ and $(1,2)$ we draw a line from $(0,1/2)$ to $(1,3/2)$, and on an E-domino in
the same position we draw a line from $(0,3/2)$ to $(1,1/2)$. Finally, on and S-domino,
placed so that it has its corners at $(0,0)$ and $(2,1)$, 
we draw a line from $(0,1/2)$ to $(2,1/2)$.
It is straightforward to see that these paths form a family of non-intersecting paths from 
$A_r=(-n-1+r,-r+1/2)$ to $B_r=(n+1-r,-r+1/2)$, $r=1,\dots, n$. The top path, from
$A_1$ to $B_1$, can be viewed as a function $t\to X_n(t)$, $|t|\le n$, and we will call it the
{\it NPR-boundary process}, since the north polar region is exactly
the part of the domino tiling that lies completely above $X_n(t)$, see fig.1.

\begin{figure}[h]
\includegraphics{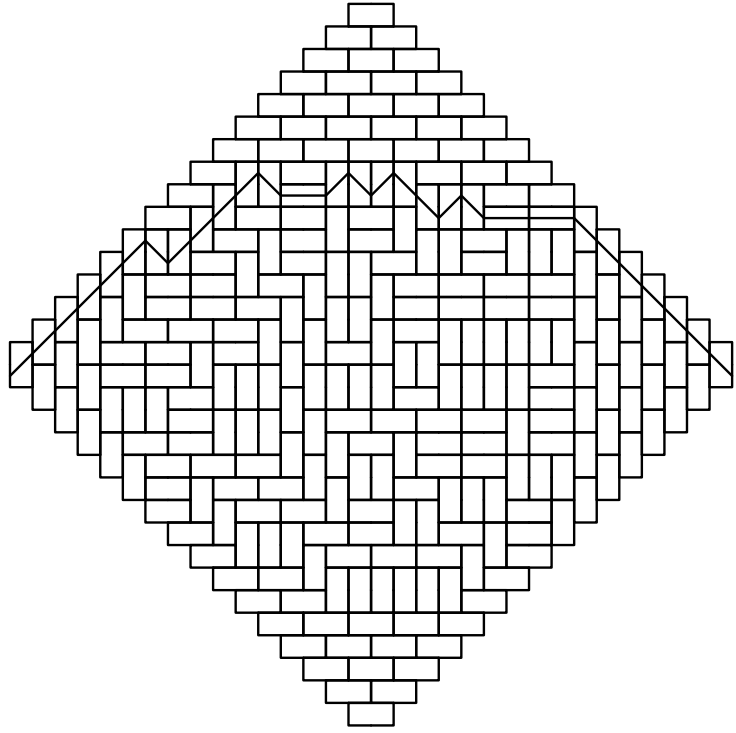}
\caption{An NPR-boundary process.}
\end{figure}

These non-intersecting paths do not immediately, using the LGV-theorem, lead to a measure
of the form (\ref{1.41}). In order to obtain a measure of this form we have to transform 
the paths. We will only outline how this is done, see \cite{JoAz} for all the details.
Introduce a new coordinate system (CS-II) with origin at $(-n,-1/2)$ and axes
$e_{II}=(1,1)$, $f_{II}=(-1,1)$ in CS-I, which gives the coordinate transformation
\begin{equation}\label{2.1}
\begin{cases}
x_1=x_2-y_2-n\\
y_1=x_2+y_2-1/2.
\end{cases}
\end{equation}
In CS-II the non-intersecting paths go from $A_j=(0,-j+1)$ to $B_j=(n+1-j,-n)$,
$1\le j\le n$, and have three types of steps $(1,0)$, $(0,-1)$ and $(1,-1)$, see fig.2. 
We can view them as non-intersecting paths in an appropriate directed graph $\mathcal{G}$.
The weight on the domino tiling can be transported to a weight on the non-intersecting
paths by letting the steps $(1,0)$, $(0,-1)$ have weight $a$ and the step $(1,-1)$ weight 1.
Take $N\ge n$ and set $A_j=(0,1-j)$ and $C_j=(n,-n+1-j)$, $1\le j\le N$, see fig. 3. 
It is not so difficult to see that if $\pi_1,\dots,\pi_N$ are non-interesecting paths from
$A_1,\dots,A_N$ to $C_1,\dots,C_N$, then $\pi_k$ has to go through $B_k$, $1\le k\le n$.
Furthermore the paths from $B_k$ to $C_k$, $1\le k\le n$, and the paths from 
$B_k$ to $C_k$, $n< k\le N$, only have steps $(1,-1)$. Hence adding the paths from $A_k$
to $C_k$, $n<k\le N$, has no effect on the correspondence with domino tilings in
the Aztec diamond or the weight, and we can just as well consider this extended system of
paths.

\begin{figure}[h]
\includegraphics{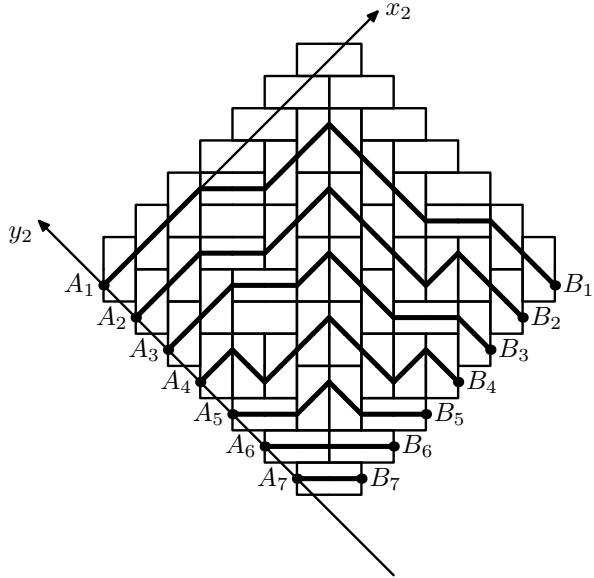}
\label{fig:csii}
\caption{CS-II and non-intersecting paths descibing the tiling.}
\end{figure}

\begin{figure}[h]
\includegraphics{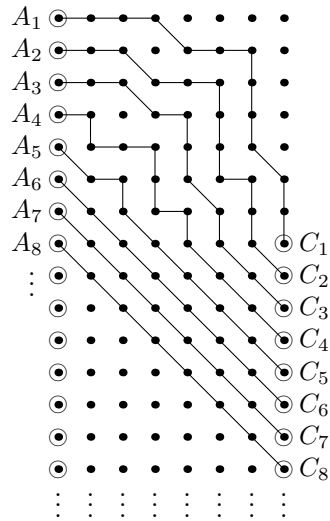}
\caption{The non-intersecting paths in the graph $\mathcal{G}$.}
\end{figure}

Each path $\pi_k$ from $A_k$ to $C_k$ has a first and a last vertex, which could coincide, 
on each vertical line
$x_2=k$, $1\le k\le n$. In order to get a measure of the type we want
we have to double the vertical lines so that the first and the last vertices on each 
vertical line ends up on different vertical lines. These first and last vertices will
form the point process we are interested in. We can also shift the paths so that the 
initial and
final points, which are fixed, end up at the same height. The result can be seen in
fig. 4. These non-intersecting paths which connect $(0,1-j)$ to $(2n,1-j)$, $1\le j\le N$, lie
in a new directed graph $\mathcal{G}'$. The steps from even to odd columns are 
$(1,0)$ with weight 1 or $(1,1)$ with weight $a$, the steps from odd to even columns
are $(1,0)$ with weight 1, and we also have steps $(0,-1)$ with weight $a$ in the even
columns. With this choice of weights we still have a weight preserving bijection with
the original domino tiling of $A_n$. The associated particles, which we think of as a point
process, are indicated in fig. 4. The NPR-boundary process corresponds to the top path
in this picture.

\begin{figure}[h!]
\includegraphics{fig.3}
\caption{The non-intersecting paths in the graph $\mathcal{G}'$ 
corresponding to the tiling in figure \ref{fig:csii}. 
The particles in the determinantal process are 
the circled dots.}
\end{figure}

The paths $\pi_1,\dots,\pi_N$ just described connecting $(0,j-1)$ to $(2n,1-j)$, 
$1\le j\le N$, 
see fig. 4, can be thought of as being built up from $2n$ transition steps. 
We have points $x^r_1, \dots,x^r_N$ on line $r$ which connect via non-intersecting paths
to points $x^{r+1}_1,\dots,x^{r+1}_N$ on line $r+1$. Let $\phi_{r,r+1}(x,y)$ be the 
transition weight to go from $x$ on the line $r$ to $y$ on the line $r+1$. It follows from 
the discussion above that
\begin{equation}\label{2.2}
\phi_{2i,2i+1}(x,y)=
\begin{cases}
\text{$a$ if $y-x=1$}\\
\text{1 if $y-x=0$}\\
\text{0 otherwise}
\end{cases}
\end{equation}
and
\begin{equation}\label{2.3}
\phi_{2i-1,2i}(x,y)=
\begin{cases}
\text{$a^{-(y-x)}$ if $y-x\le 0$}\\
\text{0 otherwise}
\end{cases}
\end{equation}
From the LGV-theorem we see that the weight of all non-intersecting paths from $x^r\in
\mathbb{Z}^n$ on line $r$ to $x^{r+1}\in\mathbb{Z}^n$ on line $r+1$ is
\begin{equation}\label{2.4}
\det(\phi_{r,r+1}(x^r_i,x^{r+1}_j))_{1\le i,j\le n}.
\end{equation}
Note that the initial and final configurations are fixed $x^0_i=1-i=x^{2n}_i$, $1\le i\le N$.
The weight of the whole configuration of non-intersecting paths is then
\begin{equation}\label{2.5}
\prod_{r=0}^{2n-1}\det(\phi_{r,r+1}(x^r_i,x^{r+1}_j))_{1\le i,j\le n}.
\end{equation}
and normalizing we obtain a probability measure of the form (\ref{1.41}). We know that the
associated point process with points $(r,x^r_j)$, $1\le r\le 2n-1$, $1\le j\le N$, has 
determinantal correlation functions with correlation kernel given by (\ref{1.44}). For reasons
that will become clear below we will call this correlation kernel the {\it
extended Krawtchouk kernel}. In order to make use of this kernel for asymptotic computations 
we have to be able to compute the inverse matrix $A^{-1}$ in (\ref{1.45}) in some way, and 
obtain a more useful formula. This will be discussed in the beginning of the section \ref{Asymptotics}.
The fluctuations of the NPR-boundary process are described by the fluctuations of the particles
$x^1_1,\dots, x^{2n-1}_1$, which are the last particles on each vertical line.
If we consider a particular line, say line $r$, then the points $x^r_1,\dots,x^r_N$ form
a determinantal point process and we can obtain the distribution function for $x^r_1$,
the last particle, from proposition \ref{Prop1.6}. We will discuss the limit theorem
that can be obtained in the section \ref{Asymptotics}.

\subsection{Relations to other models}

The north polar region can be investigated in a different way, which is related to
the corner growth model that will be studied in section 4. This is based on the so called
shuffling algorithm, \cite{EKLP}, \cite{JPS}, which is an algorithm for generating
a random tiling of $A_n$ where vertical tiles have weight $a$ and horizontal tiles
weight 1. Here is a description of the shuffling procedure following \cite{JPS}. For
a proof that it actually works, see \cite{EKLP}. The shuffling algorithm generates a
random tiling of $A_n$ starting from a random tiling of $A_{n-1}$. We can tile $A_1$ by
either two vertical dominos, with probability $q=a^2/(1+a^2)$, or two horizontal dominos,
with probability $1-q=1/(1+a^2)$. Assume now that we have generated a random tiling $T$
of $A_{n-1}$ according to the probability measure where the probability of
$T$ is proportional to $a^{v(T)}$. Two horizontal dominos sharing a side of length two
form a {\it bad} pair if the lower one is an N-domino and the upper one an S-domino, two
vertical dominos sharing a side of length two are a bad pair if the left one is an E-domino
and the right one a W-domino. Start by removing all bad pairs in $A_{n-1}$. Next, move all
remaining N-, S-, E- and W-dominos one step up, down, right and left respectively.
After these steps what remains to fill $A_n$ are $2\times 2$-blocks. In the final step
we fill each of these $2\times 2$-blocks with a vertical pair with probability $q$ and
a horizontal pair with probability $1-q$. This procedure will generate a random tiling
of $A_n$ where each vertical domino has weight $a$ and horizontal domino weight 1.
If we draw the non-intersecting paths in a somewhat different way to what was done above,
this shuffling algorithm can be translated into a certain multilayer polynuclear growth
(PNG) model, see \cite{JNIP}.

How does the north polar region evolve during the shuffling algorithm? It is clear from
the description of the algorithm that it can only grow. The growth will be directly
related to the so called {\it corner growth model} which we first define. We will return
to this model in section \ref{Corner}. Let $\lambda=(\lambda_1,\lambda_2,\lambda_3,\dots)$ be a
{\it partition}, i.e. $\lambda_i$, $i\ge 1$ are non-negative integers, $\lambda_1\ge\lambda_2
\ge\lambda_3\ge\dots$ and there is an $\ell\ge 0$ such that $\lambda_i=0$ if $i\ge \ell$.
The smallest such $\ell$ is called the {\it length}, $\ell(\lambda)$, 
of the partition. We say that $\lambda$ is a
partition of the integer $N=\lambda_1+\lambda_2+\dots$, written $\lambda\vdash N$. To the
partition $\lambda$ we associate the following set of integer points in the first 
quadrant, the {\it shape} of $\lambda$,
\begin{equation}\label{2.6}
S(\lambda)=\{(i,j)\in\mathbb{Z}_+^2\,;\,1\le i\le\lambda_j,j\ge 1\}.
\end{equation}
We can also define the {\it filled-in shape} of $\lambda$
\begin{equation}\label{2.7}
\overline{S}(\lambda)=S(\lambda)+[-1,0]^2
\end{equation}
a subset of $[0,\infty)^2$. The shape $S(\lambda)$ is one way of drawing the Young or Ferrer diagram
associated to $\lambda$. The set $\overline{S}(\lambda)\cup(\mathbb{R}^2\setminus
(0,\infty)^2)$ has  {\it corners} in certain places. These are positions where you can add 
a unit square to $\overline{S}(\lambda)$ so that it still corresponds to a partition.
Starting with the empty shape corresponding to $\lambda=(0,0,0,\dots)$ we grow 
larger shapes by adding at succesive times $1,2,\dots$ unit squares independently at each 
corner with probability $1-q$, where $0<q<1$ is fixed. We call this growth model
the {\it corner growth model}, \cite{Ro}, \cite{JPS},\cite{JoSh}. Let $S_{\text{CG}}(n)$
denote the random shape obtained at time $n$.

Consider now the evolution of the NPR under the shuffling algorithm. Put a point in the center
of each N-domino in the NPR in $A_n$. Viewed from the coordinate system (CS-III) with
origin at $(0,n+3/2)$ and axes $e_{III}=(-1,-1)$, $f_{III}=(1,-1)$ in CS-I, these points form
a random shape  $S_{\text{Az}}(n)$ of some partition $\lambda$. Analysis of the shuffling
algorithm shows that $S_{\text{CG}}(n)$ and $S_{\text{Az}}(n)$ are equal in law if $q=
a^2/(1+a^2)$ as above. Hence results for the NPR in the Aztec diamond can be translated into
results about the random shape in the corner growth model and vice-versa.

Note that the waiting time before a corner point is added is a geometric random variable starting
at 1. Let $G^\ast(M,N)$ denote the time when the point $(M,N)\in\mathbb{Z}_+^2$ is added
to the square. It follows from the corner geometry that before the point $(M,N)$ is added
to the shape the points $(M-1,N)$ and $(M,N-1)$ must already have been added. Let
$w(i,j)$, $i,j\ge 1$ denote independent geometric random variables starting at 0,
\begin{equation}\label{2.8}
\mathbb{P}[w(i,j)=m]=(1-q)q^m,
\end{equation}
$m\ge 0$, and define recursively
\begin{equation}\label{2.9}
G(M,N)=\max (G(M-1,N), G(M,N-1))+w(M,N).
\end{equation}
It follows from the lack of memory property of the geometric distribution that
\begin{equation}\label{2.10}
G^\ast(M,N)=G(M,N)+M+N-1.
\end{equation}
Consequently, we have
\begin{equation}\label{2.11}
S_{\text{CG}}(n)=\{(M,N)\in\mathbb{Z}_+^2\,;\,G(M,N)+M+N-1\le n\}.
\end{equation}
There is a translation of the growth rule (\ref{2.9}) into a PNG-type model, see
\cite{PrSp}, \cite{JNIP}, \cite{JDPNG}. This polynuclear growth model is not the same
as the one coming from the shuffling procedure above.

There is also a translation of the corner growth model into a totally asymmetric simple
exclusion process (TASEP) in discrete time, \cite{Ro}, \cite{JPS}, \cite{JoSh}. 
This connection
was used in \cite{JPS} to prove the so called arctic circle theorem. The complement of the
four polar regions in a tiling of the Aztec diamond is called the {\it temperate region},
and is the region where the tiling is more or less random. The {\it arctic circle theorem}
says that, in the case of a uniform random tiling of $A_n$, the boundary of the
temperate region scaled
down by $n$, converges almost surely to a circle. When the weight $a\neq 1$ we get an ellipse 
instead. This can be translated to a limit of $G(M,N)/N$ as $M,N\to\infty$, $M/N\to\text{const}
>0$ ({\it time constant}). In section \ref{Asymptotics} we will see how the representation of the random
tiling as a determinantal process allows us to get precise information about the fluctuations
of the arctic ellipse or, equivalently, of the NPR-boundary process. This then also leads
to precise limit theorems for $G(M,N)$.

The quantity $G(M,N)$ also has a {\it last-passage time} interpretation, which follows
from the recursion (\ref{2.9}). An {\it up/right path} $\pi$ from $(1,1)$ to $(M,N)$ is a
sequence of points $(i_k,j_k)$, $0\le k\le M+N-1$, such that $(i_0,j_0)=(1,1)$, 
$(i_{M+N-1}, j_{M+N-1})=(M,N)$ and $(i_{k+1},j_{k+1})-(i_k,j_k)=(1,0)$ or $(0,1)$. Viewed
from a coordinate system rotated $45^\circ$ it is a simple random walk path. We have that
\begin{equation}\label{2.12}
G(M,N)=\max_{\pi}\sum_{(i,j)\in\pi} w(i,j),
\end{equation}
where the maximum is over all up/right paths from $(1,1)$ to $(M,N)$.

There is an interesting limit of $G(N,N)$ as $N\to\infty$ if we choose the parameter
$q=\alpha/N^2$. With this choice of $q$ it is not hard to see, \cite{JDOPE}, that among
the $w(i,j)$, $1\le i,j\le N$, in each row and column there will be at most a single 1,
and in the whole square all numbers will be $\le 1$ with probability $\to 1$ as $N\to
\infty$. In the limit $N\to\infty$ the set $\{(i,j)\,;\,1\le i,j\le N\,,\,w(i,j)\ge 1\}$
scaled down by $N$ will converge to a Poisson process in $[0,1]^2$ with intensity
$\alpha$. The number $n$ of points in $[0,1]^2$ is a Po($\alpha$) random variable. Let
$x_1<\dots<x_n$ and $y_1<\dots y_n$ be the $x$- and $y$-coordinates of these points.
The coordinates of the points in the Poisson process can then be written
$(x_j,y_{\sigma(j)})$, $j=1,\dots,n$, where $\sigma\in S_n$ is a permutation of 
$\{1,\dots, n\}$. The uniformity of the Poisson process implies that $\sigma$ will be a uniform
random permutation from $S_n$. Denote by $L(\alpha)$ the length of the {\it longest increasing
subsequence} in this permutation. We say that $\sigma(i_1),\dots,\sigma(i_\ell)$ is
an increasing subsequence in $\sigma$ if $i_1<\dots<i_\ell$ and $\sigma(i_1)<\dots<
\sigma(i_\ell)$. If we look back at (\ref{2.11}) we see that we should have
\begin{equation}\label{2.13}
G(N,N)\to L(\alpha)
\end{equation}
in distribution as $N\to\infty$. Hence, we may be able to use 
results for random tilings of the Aztec diamond also to study the problem of the distribution
of the length of the longest increasing subsequence in a random permutation.

\section{Asymptotics}\label{Asymptotics}

\subsection{Double contour integral formula for the correlation kernel}

In order for the formulas (\ref{1.44}) and (\ref{1.45}) for the correlation kernel of the
determinantal process defined by (\ref{1.41}) to be useful we have to find a different
representation. In particular we need some way of computing the inverse matrix $A^{-1}$.
When $A$ is a Toeplitz matrix it may be possible to do this, at least approximately,
by using a Wiener-Hopf factorization of the symbol for $A$. For the models that we will consider
$A$ actually is a Toeplitz matrix and we will be able to find nice formulas.

The space $X$ in (\ref{1.41}) will now be $\mathbb{Z}$, and we will also take
$X_0=X_{m+1}$. Hence $\underline{x}\in\mathbb({Z}^n)^m$.
Let $f_r(z)$, $z=e^{i\theta}$, be a function in $L^1(\mathbb{T})$ with Fourier coefficients
$\hat{f}_r(n)$, $n\in\mathbb{Z}$. Assume that the transition weights $\phi_{r,r+1}$ in
(\ref{1.41}) are given by
\begin{equation}\label{3.1}
\phi_{r,r+1}(x,y)=\hat{f}_r(y-x),
\end{equation}
$0\le r\le m$, $x,y\in\mathbb{Z}$. Then, for $r<s$,
\begin{equation}\label{3.2}
\phi_{r,s}(x,y)=\hat{f}_{r,s}(y-x),
\end{equation}
where
\begin{equation}\label{3.3}
f_{r,s}(z)=\prod_{\ell=r}^{s-1}f_\ell(z).
\end{equation}
We see that $a_{ij}=\hat{f}_{0,m+1}(x^{m+1}_j-x^0_i)=\hat{f}_{0,m+1}(i-j)$, since 
$x^{m+1}_j=x^0_j=1-j$. The matrix $A$ is thus a Toeplitz matrix with symbol
$a(z)=f_{0,m+1}(z)$, $A=T_n(a)$, where $T_n(a)=(\hat{a}(i-j))_{1\le i,j\le n}$.
A computation, \cite{JDPNG}, shows that
\begin{equation}\label{3.4}
\sum_{x,y\in\mathbb{Z}}\tilde{K}_{n,m}(r,x;s,y)z^xw^{-y}=
\frac zw f_{r,n+1}(\frac 1z)f_{0,s}(\frac 1w)\sum_{i,j=1}^n z^{-i}(T_n^{-1}(a))_{ij}w^j.
\end{equation}

Let $T(a)=(\hat{a}(i-j))_{i,j\ge 1}$ denote the infinite Toeplitz matrix with symbol $a$.
We say that $a\in L^1(\mathbb{T})$ 
has a Wiener-Hopf factorization, if it can be written $a=a_+a_-$ on $\mathbb{T}$, where
$a^+(z)=\sum_{n=0}^\infty a_n^+ z^n$, $a^-(z)=\sum_{n=0}^\infty a_n^- z^{-n}$ ,
$(a^+_n), (a^-_n)\in\ell^1$. We can extend $a_+$ to $|z|\le 1$ and $a_-$ to 
$\{|z|\ge 1\}\cup\{\infty\}$. We also require 
that $a_+$ and $a_-$ have no zeros in these regions
and that $a$ has winding number zero with respect to the origin. Also, suppose that
\begin{equation}\label{3.4'}
\sum_{n\in\mathbb{Z}} |n|^\alpha |\hat{a}_n|<\infty,
\end{equation}
for some $\alpha>0$. Then, $T_n(a)$ is invertible for $n$ sufficiently large and
\begin{equation}\label{3.5}
\left|(T_n^{-1}(a))_{jk}-(T(a_+^{-1})T(a_-^{-1}))_{jk}\right|
\le C\min ((n+1-k)^{-\alpha},(n+1-j)^{-\alpha} )
\end{equation}
for some constant $C$, $1\le j,k\le n$.

If $f_r$ is analytic in $1-\epsilon<|z|<1+\epsilon$ for some $\epsilon>0$, 
and has a Wiener-Hopf factorization as defined above, then $a(z)=f_{0,m+1}(z)$
will have a Wiener-Hopf factorization and (\ref{3.4'}) will be satisfied. Hence we can
use (\ref{3.5}) to compute the inverse of the Toeplitz matrix. Combined with 
(\ref{3.4}) this yields
\begin{equation}\label{3.6}
\lim_{n\to\infty}\sum_{x,y\in\mathbb{Z}}\tilde{K}_{n,m}(r,x;s,y)z^xw^{-y}=
\frac z{z-w} F(z,w),
\end{equation}
for $1-\epsilon<|w|<1<|z|<1+\epsilon$, where
\begin{equation}\label{3.7}
F(z,w)=\prod_{t=r}^m f^-_t(\frac 1z)\prod_{t=0}^{s-1} f^+_t(\frac 1w)
\prod_{t=0}^{r-1} f^+_t(\frac 1z)^{-1}\prod_{t=s}^m f^-_t(\frac 1w)^{-1}.
\end{equation}
Hence,
\begin{align}\label{3.8}
&\lim_{n\to\infty}\tilde{K}_{n,m}(r,x;s,y)=\tilde{K}_{m}(r,x;s,y)
\notag\\
&\doteq\frac 1{(2\pi i)^2}\int_{\gamma_{r_2}}\frac {dz}z\int_{\gamma_{r_1}}\frac {dw}w
\frac{w^y}{z^x}\frac{z}{z-w} F(z,w),
\end{align}
where $\gamma_r$ is a circle with radius $r$ and center $0$, $1-\epsilon<r_1<r_2<1+
\epsilon$. We obtain
\begin{equation}
\lim_{n\to\infty} K_{n,m}(r,x;s,y)=-\phi_{r,s}(x,y)+\tilde{K}_{m}(r,x;s,y)
\doteq K_{m}(r,x;s,y).
\end{equation}
From (\ref{3.2}) and (\ref{3.7}) it follows that
\begin{equation}\label{3.9}
\phi_{r,s}(x,y)=\frac 1{2\pi}\int_{-\pi}^\pi e^{i(y-x)\theta} F(e^{i\theta},e^{i\theta})
d\theta.
\end{equation}
The integral formula (\ref{3.8}) gives $K_{m}(r,x;s,y)$ if $r\ge s$. Using (\ref{3.9})
and the residue theorem we see that it also gives $K_{m}(r,x;s,y)$ for $r<s$ if we
take $1-\epsilon<r_2<r_1<1+\epsilon$. These integral formulas are good representations of the 
correlation kernel if we want to investigate its asymptotics.

\subsection{Asymptotics for the Aztec diamond}

In our discussion of the Aztec diamond we saw that adding more paths from $A_k$ to $C_k$ for
$k>n$ had no effect on the probability measure since all these paths are fixed. Hence, we
can take the limit $N\to\infty$ in the correlation kernel $K_{N,n}(r,x;s,y)$ coming from
the measure (\ref{2.5}) with transition functions (\ref{2.2}), (\ref{2.3}), without changing
anything. We get a determinantal process on $2n-1$ copies of $\mathbb{Z}$ where the 
configuration is always frozen below the level $-(n-1)$. We will thus be able to use the
formula (\ref{3.8}). Assume that $0< a<1$. The case
$a=1$ can be handled by considering the limit $a\to 1$ and using continuity. Set
$f_{2i}(z)=az+1$ and $f_{2i+1}(z)=(1-a/z)^{-1}$. Then (\ref{3.1}) holds and all the
conditions on $f_r$ above are satisfied, $f_{2i}^+(z)=az+1$, $f_{2i}^-(z)=f_{2i+1}^+(z)=1$
and $f_{2i+1}^-(z)=(1-a/z)^{-1}$. Denote the generating function in (\ref{3.7}) by
$F_{n,r,s}(z,w)$ to indicate the dependence on $n,r,s$, $0<r,s<2n$.
We obtain,
\begin{equation}\label{3.10}
F_{n,2r-\epsilon_1,2s-\epsilon_2}(z,w)=\frac{(1-aw)^{n-s+\epsilon_2}
(1+a/w)^s}{(1-az)^{n-r+\epsilon_1}(1+a/z)^r},
\end{equation}
with $\epsilon_i\in\{0,1\}$.
The Aztec diamond particle process is a determintal process with kernel
\begin{equation}\label{3.11}
K_{\text{Kr},n}(r,x;s,y)=-\phi_{r,s}(x,y)+
\frac 1{(2\pi i)^2}\int_{\gamma_{r_2}}\frac {dz}z\int_{\gamma_{r_1}}\frac {dw}w
\frac{w^y}{z^x}\frac{z}{z-w} F_{n,r,s}(z,w),
\end{equation}
where $a<r_2<1/a$, $0<r_1<r_2$. Here, by (\ref{3.9}),
\begin{equation}\label{3.12}
\phi_{r,s}(x,y)=
\frac 1{(2\pi i)}\int_{\gamma_{r}}\frac {dz}z
z^{y-x}F_{n,r,s}(z,z)
\end{equation}
We will call the kernel (\ref{3.11}) the {\it extended Krawtchouk kernel}. The reason for this
name is that it can be expressed in terms of Krawtchouk polynomials, see \cite{JoAz}
for all the details. Here we will only consider the case $r=s$. Let $p_k(x;q,n)$ be the
normalized {\it Krawtchouk polynomial}, i.e. it is a polynomial of degree $k$ satisfying
the orthogonality condition
\begin{equation}
\sum_{x=0}^np_j(x;q,n)p_k(x;q,n)\binom nx q^x(1-q)^{n-x}=\delta_{jk},
\notag
\end{equation}
$0\le j,k\le n$, on $\{0,\dots,n\}$. 
Define the {\it Krawtchouk kernel},
\begin{equation}\label{3.13}
K_{\text{Kr},n,r,q}(x,y)=\sum_{k=0}^{r-1}p_k(x;q,n)p_k(y;q,n)
\left[\binom nx q^x(1-q)^{n-x}\binom ny q^y(1-q)^{n-y}\right]^{1/2}.
\end{equation}
It can then be shown, \cite{JoAz}, using the contour
integral formula for $p_j(x;q,n)$ that
\begin{equation}\label{3.14}
K_{\text{Kr},n}(2(n-r)+1,x-r+1;2(n-r)+1,y-r+1)=K_{\text{Kr},n,r,q}(x,y),
\end{equation}
where $q=a^2/(1+a^2)$.

Note the similarity between the Krawtchouk kernel and the Hermite kernel (\ref{1.40}).
The same argument that showed that the point process defined by (\ref{1.20}) has determinantal 
correlation functions with kernel (\ref{1.40}) shows that the probability measure,
the {\it Krawtchouk ensemble},
\begin{equation}\label{3.15}
u_{n,r}(x_1,\dots,x_r)=\frac 1{Z_{n,r}}\prod_{1\le i<j\le r}(x_i-x_j)^2\prod_{j=1}^n
\binom{n}{x_j}q^{x_j}(1-q)^{n-x_j}
\end{equation}
on $\{0,\dots,n\}^r$ defines a determinantal process on $\{0,\dots,n\}$ with correlation
kernel (\ref{3.13}). Probability measures of the form (\ref{1.20}) and (\ref{3.15}),
i.e. a Vandermonde determinant squared times a product of single particle weights are
called {\it orthogonal polynomial ensembles} and go back to the early work of Gaudin and
Mehta, \cite{Me}. The Krawtchouk ensemble is an example of a discrete orthogonal
polynomial ensemble, \cite{JDOPE}.

The NPR-boundary process is related to the top particles $x^r_{\max}=\max x^r_j$ in the
Aztec diamond particle process. If we look first at a single line $r$ we see that $x^r_{\max}$
is the last particle in a point process on $\mathbb{Z}$ given by the $x^r_i$. It follows from
proposition \ref{Prop1.6} and (\ref{3.14}) that
\begin{equation}\label{3.16}
\mathbb{P}[x^{2(n-r)+1}_{\max}\le t-r+1]=\det(I-K_{\text{Kr},r,n,q}\chi_{(t,\infty)})_
{\ell^2(\mathbb{Z})}
\end{equation}
for any $t\in\mathbb{R}$. Looking at the geometry of the Aztec diamond and how the 
non-intersecting paths were defined we see that the NPR-boundary process $X_n(t)$ is obtained
by joining the points (in CS-I)
$Q(j)=(2j-x^{2j}_{\max}-n,x^{2j}_{\max}-1/2)$, 
$P(j)=(2j-x^{2j}_{\max}-n,x^{2j-1}_{\max}-1/2)$,
$1\le j\le n$ with straight lines. Hence, (\ref{3.16}) can be used to investigate $X_n(t)$.

We see that asymptotics for the Krawtchouk kernel will give us asymptotics for the
NPR-boundary process. The relation between the NPR-boundary process and the corner
growth model discussed above yields
\begin{equation}\label{3.17}
\mathbb{P}[G^\ast(M,N)\le n]=\mathbb{P}_{\text{Kr},n}[x^{2(n-M)+1}_{\max}\le n+1-M-N].
\end{equation}
Combining (\ref{3.16}) and (\ref{3.17}) and using (\ref{2.10}) we find
\begin{equation}\label{3.18}
\mathbb{P}[G(M,N)\le t]=\det(I-K_{\text{Kr},M,t+M+N-1,q})_{\ell^2(\{t+M+1,t+M+2,\dots\})}
\end{equation}
$t\in\mathbb{Z}$, see also \cite{JNIP}.

We will not discuss the asymptotic analysis in any detail, but only give the results and 
outline the main structure of the proofs. From (\ref{3.11}), (\ref{3.14}) and Cauchy's 
theorem we see that the Krawtchouk kernel has the representation
\begin{equation}\label{3.19}
K_{\text{Kr},r,n,q}(x,y)=\frac 1{(2\pi i)^2}\int_{\Gamma}\frac {dz}z\int_\gamma\frac{dw}w 
\frac {z}{z-w}\frac{z^{n-x}}{w^{n-y}}\frac{(1-aw)^r(w+a)^{n-r+1}}{(1-az)^r(z+a)^{n-r+1}},
\end{equation}
where $\Gamma$ is given by $t\to\alpha_2+it$, $t\in\mathbb{R}$ and $\gamma$ is a circle
with radus $\alpha_1$ centered at the origin, $0<\alpha_1<\alpha_2<1/a$.

We want to show that close to the rightmost particle the Krawtchouk kernel, appropriately
scaled, converges to the Airy kernel given by (\ref{1.28}) so that in the scaling limit 
close to the edge of the Krawtchouk ensemble we get the Airy kernel point process.
The last particle in the Krawtchouk ensemble, and hence the last particle in
the Aztec diamond point process restricted to a line, will in the limit fluctuate
according to the Tracy-Widom distribution. In order to show this it is useful to have
a double contour integral representation of the Airy kernel. Using the formula
\begin{equation}\label{3.20}
\Ai(x)=\frac 1{2\pi}\int_{\im z=\eta}e^{iz^3/3+ixz}dz,
\end{equation}
with $\eta>0$, for the Airy function in (\ref{1.28}) we obtain
\begin{equation}\label{3.21}
A(x,y)=\frac 1{(2\pi i)^2}\int_{\im z=\eta}dz\int_{\im w=\eta}dw
\frac{e^{ixz+iyw+i(z^3+w^3)/3}}
{i(z+w)}.
\end{equation}
In discussing asymptotic results for the Krawtchouk kernel we will for simplicity
only consider the case $a=1$ and the part of the NPR-boundary that lies above 
a neighbourhood of $x_1=0$.

\begin{theorem}\label{Thm3.1}
Set $\beta=2^{-3/2}(\sqrt{2}+1)$ and $\gamma=2^{-3/2}(\sqrt{2}-1)$. If $r=\gamma n$,
$x=\beta n+2^{-5/6}n^{1/3}\xi$ and $y=\beta n+2^{-5/6}n^{1/3}\eta$, then
\begin{equation}\label{3.22}
2^{-5/6}n^{1/3}(\sqrt{2}-1)^{x-y}K_{\text{Kr},r,n,1/2}(x,y)\to A(\xi,\eta)
\end{equation}
as $n\to\infty$, uniformly for $\xi,\eta$ in a compact set in $\mathbb{R}$. 
Also, if $X_{\max}$ denotes the rightmost particle in the Krawtchouk ensemble with $q=1/2$, 
$r=\gamma n$, then
\begin{equation}\label{3.23}
\mathbb{P}[X_{\max}\le \beta n+2^{-5/6}n^{1/3}\xi]\to F_{\text{TW}}(\xi)
\end{equation}
as $N\to\infty$ for every $\xi$.
\end{theorem}

\begin{proof} (Sketch) \cite{JoAz}. The integral (\ref{3.19}) can be written
($a=1$, which corresponds to $q=1/2$),
\begin{equation}\label{3.24}
K_{\text{Kr},r,n,1/2}(x,y)=\frac 1{(2\pi i)^2}\int_{\Gamma}\frac {dz}z\int_\gamma\frac{dw}w 
\frac {z}{z-w}e^{nf(z)-nf(w)}\frac{g(z)}{g(w)},
\end{equation}
where $e^{nf(z)}$ represents the leading order behaviour of the integrand and
\begin{equation}
f(z)=(1-\beta)\log z-\gamma\log (1-z)-(1-\gamma)\log(z+1).
\notag
\end{equation}
We want to apply a steepest descent argument to this integral. The saddle point 
condition $f'(z)=0$ gives $\beta z^2+(2\gamma-1)z+1-\beta=0$, which has a double root
if $(\gamma-1/2)^2+\beta^2-\beta=0$. This can also be written,
\begin{equation}\label{3.25}
(1-\beta-\gamma)^2+(\beta-\gamma)^2=1/2.
\end{equation}
We see that the $\beta,\gamma$ in the theorem satisfy (\ref{3.25}) and hence we have an 
Airy-type steepest descent problem in (\ref{3.24}). If we carry out the argument in detail 
we find that
\begin{align}
&\frac{2^{-5/6}n^{1/3}(\sqrt{2}-1)^{x-y}}{(2\pi i)^2}
\int_{\Gamma}\frac {dz}z\int_\gamma\frac{dw}w 
\frac {z}{z-w}e^{nf(z)-nf(w)}\frac{g(z)}{g(w)}
\notag\\
&\to\frac 1{(2\pi i)^2}\int_{\im z=\eta}dz\int_{\im w=\eta}dw\frac{e^{ixz+iyw+i(z^3+w^3)/3}}
{i(z+w)}
\notag
\end{align}
as $N\to\infty$. Together with (\ref{3.21}) this gives (\ref{3.22}).

We can also use (\ref{3.24}) to derive estimates of the Krawtchouk kernel. By proposition
\ref{Prop1.6}, the left hand side of (\ref{3.23}) can be written as a Fredholm expansion, 
and (\ref{3.22}) toghether with the estimates can be used to show that this Fredholm
expansion converges to the Fredholm expansion for the TW-distribution (\ref{1.29}).
\end{proof}

The equation (\ref{3.25}) is actually the equation for the arctic circle. If we look at
the relation (\ref{3.16}) between $x^{2(n-r)+1}_{\max}$ and the position of the last particle
in the Krawtchouk ensemble, and translate this back to our original coordinate system
(CS-I) we obtain $x_1^2+y_1^2=1/2$, which is the equation for the arctic circle. If we
instead of $a=1$ we took $0<a<1$, the equation for the arctic ellipse is
$x_1^2/p+y_1^2/q=1$, $q=a^2/(1+a^2)$, $p=1-q$, which can be obtained similarly.

From (\ref{3.16}) and the relation between $x^r_{\max}$ and the NPR-boundary process we 
obtain
\begin{equation}\label{3.26}
\mathbb{P}[X_n(0)\le n/\sqrt{2}+2^{-5/6} n^{1/3}\xi]\to F_{\text{TW}}(\xi)
\end{equation}
as $n\to\infty$. We will consider a generalization of (\ref{3.26}) in theorem
\ref{Thm3.3}.

From (\ref{3.24}) and (\ref{3.18}) (up to some technical details) we also obtain, in the case 
$q=1/2$,
\begin{equation}\label{3.27}
\lim_{N\to\infty}\mathbb{P}[G(N,N)\le 2(\sqrt{2}+1)N+2^{1/6}(\sqrt{2}+1)^{4/3}N^{1/3}\xi]=
F_{\text{TW}}(\xi).
\end{equation}
We will discuss a more general result in section \ref{Corner}. 
The full result theorem \ref{Thm4.3} can
also be proved starting from (\ref{3.18}), (\ref{3.19}) and (\ref{3.21}).

Since we have the formula (\ref{3.11}) for the extended Krawtchouk kernel we should also
be able to derive simultaneous distributions for the last particles $x^r_{\max}$ for several
lines, and hence study the convergence of the NPR-boundary process to a limiting
stochastic process. Using (\ref{3.14}) the limit (\ref{3.22}) can be can be translated into
a limit formula for $K_{\text{Kr},n}(2r+1,x;2s+1,y)$ when $r=s$. We now  want to generalize
this to the case $r\neq s$. The limiting correlation kernel will be the so called
{\it extended Airy kernel}. It is defined by
\begin{equation}\label{3.28}
A(\tau,\xi;\sigma,\eta)=
\begin{cases}
\int_0^\infty e^{-\lambda(\tau-\sigma)}\Ai(\xi+\lambda)\Ai(\eta+\lambda)
d\lambda, &\text{if $\tau\ge\sigma$}\\
-\int_{-\infty}^0 e^{-\lambda(\tau-\sigma)}\Ai(\xi+\lambda)\Ai(\eta+\lambda)
d\lambda,  &\text{if $\tau<\sigma$.}
\end{cases}
\end{equation}
Note that $A(\tau,\xi;\tau,\eta)=A(\xi,\eta)$. Using (\ref{3.20}) it is again possible to 
rewrite this as a double contour integral. In fact,
\begin{equation}\label{3.29}
A(\tau,x;\sigma;y)=
\frac 1{(2\pi i)^2}\int_{\im z=\eta}dz\int_{\im w=\eta}dw\frac{e^{ixz+iyw+i(z^3+w^3)/3}}
{\sigma-\tau+i(z+w)}.
\end{equation}
where $\eta>0$ and $2\eta+\tau-\sigma<0$ in the case $\sigma>\tau$.

It can be shown, \cite{PrSp}, \cite{JDPNG}, that there is a stochastic process, 
the {\it Airy process}, $\tau\to \mathcal{A}(\tau)$ with continuous sample paths 
almost surely such that
\begin{equation}\label{3.30}
\mathbb{P}[\mathcal{A}(\tau_1)\le\xi_1,\dots,\mathcal{A}(\tau_m)\le\xi_m]
=\det(I-fAf)_{L^2(\{\tau_1,\dots,\tau_m\}\times\mathbb{R})},
\end{equation}
where $A$ is the extended Airy kernel and $f(\tau_j,x)=\chi_{(\xi_j,\infty)}(x)$,
$1\le j\le m$.

We can now show the following theorem. The proof is similar to that of theorem
\ref{Thm3.1} but somewhat more involved. It is based on (\ref{3.11}) and
(\ref{3.29}), see \cite{JoAz}.

\begin{theorem}\label{Thm3.2}
Define the rescaled variables $\xi,\eta,\tau,\sigma$ by $2r=n(1+1/\sqrt{2})+2^{-1/6}\tau
n^{2/3}$, $2s=n(1+1/\sqrt{2})+2^{-1/6}\sigma n^{2/3}$, $x=n/\sqrt{2}+2^{-5/6}(\xi-\tau^2)
n^{1/3}$ and $y=n/\sqrt{2}+2^{-5/6}(\eta-\tau^2)n^{1/3}$. Take $a=1$. Then,
\begin{equation}\label{3.31}
\lim_{n\to\infty} (\sqrt{2})^{x-y+2(s-r)} e^{\xi\tau-\eta\sigma-\tau^3/3+\sigma^3/3}
K_{\text{Kr},n}(2r,x;2s,y)=A(\tau,\xi;\sigma,\eta),
\end{equation}
uniformly for $\xi,\eta,\tau,\sigma$ in compact sets.
\end{theorem}

We can now give a theorem that says that the appropriately rescaled NPR-boundary process 
converges to the Airy process.

\begin{theorem}\label{Thm3.3}
Let $X_n(t)$ be the NPR-boundary process and $\mathcal{A}(\tau)$ the Airy process,
and let the weight $a$ for vertical dominos $=1$, so that we have a uniform random
tiling of the Aztec diamond. Then,
\begin{equation}\label{3.32}
\frac{X_n(2^{-1/6}n^{2/3}\tau)-n/\sqrt{2}}{2^{-5/6}n^{1/3}}
\to\mathcal{A}(\tau)-\tau^2,
\end{equation}
as $n\to\infty$, in the sense of convergence of finite-dimensional distributions.
\end{theorem}

\begin{proof}
(Sketch), \cite{JoAz}. The joint distribution of the left hand side of (\ref{3.32})
for different times $\tau$ can be expressed in terms of the joint distribution of
$\max_{k\ge 1} x^r_k$ with appropriate $r$. The fact that $x^r_k$ form a determinantal
process whose kernel is the extended Krawtchouk kernel gives that the joint 
distribution is a Fredholm determinant involving this kernel. The limit (\ref{3.31})
and some estimates can be used to show that this Fredholm determinant converges to
a Fredholm determinant like (\ref{3.30}) involving the extended Airy kernel. This 
Fredholm determinant will give the joint distribution of the right hand side of (\ref{3.32}).
\end{proof}

\subsection{Asymptotics for random permutations}
As above we let $L(\alpha)$ denote the length of the longest increasing subsequence
in a uniform random permutation from $S_N$ where $N$ is an independent 
Poisson($\alpha$)  random
variable. From (\ref{2.13}) we know that $G(N,N)\to L(\alpha)$ in distribution
as $N\to\infty$ if $q=\alpha/N^2$. Thus by (\ref{3.18}),
\begin{align}\label{3.33}
&\mathbb{P}[L(\alpha)\le n]=\lim_{N\to\infty} \mathbb{P}[G(N,N)\le n]
\notag\\
&=\lim_{N\to\infty}\det (I-K_{\text{Kr}\,,N,n+2N-1,\alpha/N^2})_{\ell^2(\{n+N,n+N+1,\dots\})}.
\end{align}
Let $a_N$ be given by $a_N^2/(1+a_N^2)=\alpha/N^2$, so that essentially $a_N=\sqrt{\alpha}
/N$. Then, by (\ref{3.19},
\begin{align}
&K_{\text{Kr}\,,N,n+2N-1,\alpha/N^2}(x+N,y+N)
\notag\\
&=\frac 1{(2\pi i)^2}\int_{\gamma_{r_2}}\frac{dz}z\int_{\gamma_{r_1}}\frac{dw}w
\frac z{z-w}\frac{w^{y+1}}{z^{x+1}}\frac{(1-a_Nw)^N(1+a_N/w)^{N+n}}
{(1-a_Nz)^N(1+a_N/z)^{N+n}},
\notag
\end{align}
where $a_N<r_1<r_2<1/a_N$. Here we can let $N\to\infty$ and obtain 
\begin{align}\label{3.34}
&\lim_{N\to\infty}K_{\text{Kr}\,,N,n+2N-1,\alpha/N^2}(x+N,y+N)
\notag\\
&=\frac 1{(2\pi i)^2}\int_{\gamma_{r_2}}\frac{dz}z\int_{\gamma_{r_1}}\frac{dw}w
\frac 1{1-w/z}\frac{w^{y+1}}{z^{x+1}}e^{-\sqrt{\alpha}(w-1/w)+\sqrt{\alpha}(z-1/z)}
\notag\\
&=\sum_{k=0}^\infty\left(\frac 1{2\pi}\int_{-\pi}^\pi e^{i(k+y+1)\theta-
2\sqrt{\alpha} i\sin\theta}d\theta\right)
\left(\frac 1{2\pi}\int_{-\pi}^\pi e^{-i(k+x+1)\theta+
2\sqrt{\alpha} i\sin\theta}d\theta\right)
\notag\\
&=\sum_{k=1}^\infty J_{x+k}(2\sqrt{\alpha})J_{y+k}(2\sqrt{\alpha})\doteq B^\alpha(x,y).
\end{align}
The second equality follows by expanding $(1-w/z)^{-1}$ in a geometric series. The
kernel $B^\alpha(x,y)$ on $\ell^2(\mathbb{Z})$ is called the {\it discrete Bessel kernel}.
The limits (\ref{3.33}) and (\ref{3.34}) and some estimates of the Krawtchouk
and discrete Bessel kernels now yield the following theorem, \cite{BOO}, \cite{JDOPE}.

\begin{theorem}\label{Thm3.4}
Let $N$ be a Poisson$(\alpha)$ random variable and pick independently a permutation
$\sigma$ from $S_N$ with the uniform distribution. Denote by $L(\alpha)$ the length
of the longest increasing subsequence in $\sigma$. Then,
\begin{equation}\label{3.35}
\mathbb{P}[L(\alpha)\le n]=\det(I-B^\alpha)_{\ell^2(\{n,n+1,\dots\})}.
\end{equation}
\end{theorem}

The formula (\ref{3.35}) can be used to prove a limit theorem for $L(\alpha)$. This theorem
was first proved in \cite{BDJ} using a completely different approach.

\begin{theorem}\label{Thm3.5}
With $L(\alpha)$ as in the previous theorem we have
\begin{equation}\label{3.36}
\mathbb{P}[\frac{L(\alpha)-2\sqrt{\alpha}}{\alpha^{1/6}}\le t]
=F_{\text{TW}}(t)
\end{equation}
as $\alpha\to\infty$.
\end{theorem}

\begin{proof} (Sketch). The Bessel functions have the following asymptotics
\begin{equation}\label{3.37}
\alpha^{1/6}J_{2\sqrt{\alpha}+\xi\alpha^{1/6}}(2\sqrt{\alpha})\to\Ai(\xi)
\end{equation}
uniformly for $\xi$ in a compact interval as $\alpha\to\infty$. This, together
with appropriate estimates, gives
\begin{equation}
\alpha^{1/6}B^\alpha(2\sqrt{\alpha}+\xi\alpha^{1/6},2\sqrt{\alpha}+\eta\alpha^{1/6})
\to\int_0^\infty\Ai(\xi+t)\Ai(\eta+t)dt=A(\xi,\eta)
\end{equation}
as $\alpha\to\infty$. Thus, by (\ref{3.35}),
\begin{equation}
\lim_{\alpha\to\infty}\mathbb{P}[\frac{L(\alpha)-2\sqrt{\alpha}}{\alpha^{1/6}}\le t]
=\det(I-A)_{L^2(t,\infty)}=F_{\text{TW}}(t).
\end{equation}
\end{proof}

If $\ell_N(\sigma)$ denotes the length of the longest increasing subsequence in a 
uniform random permutation from $S_N$, a de-Poissonization argument, \cite{Jo1},
\cite{BDJ}, makes it possible to deduce
\begin{equation}
\lim_{N\to\infty}\mathbb{P}[\frac{\ell_N(\sigma)-2\sqrt{N}}{N^{1/6}}\le t]
=F_{\text{TW}}(t),
\end{equation}
from the previous theorem.

\section{The corner growth model}\label{Corner}

\subsection{Mapping to non-intersecting paths}

In this section we will give another approach to the corner growth model by mapping
it to non-intersecting paths in a different way than that related to the Aztec
diamond.

Consider a {\it right/down path} $\lambda$ from $(0,L)$ to
$(K,0)$, i.e. a sequence of points in $\mathbb{Z}^2$, $P_j(\lambda)=(x_j,y_j)$, $j=0,
\dots,K+L$, such that $P_0(\lambda)=(0,L)$, $P_{K+L}(\lambda)=(K,0)$ and 
$P_{j+1}(\lambda)-P_{j}(\lambda)=(1,0)$ or $(0,-1)$. We use the same notation, $\lambda$, 
as for a partition since there is a unique associated partition, $\lambda_k=
\max\{x_j\,;\,y_j=k-1\}$. Note that $\lambda_1=K$ and $\ell(\lambda)=L$.

Let $w(i,j)$, $(i,j)\in\mathbb{Z}_+^2$, be independent geometric random variables with
parameter $a_ib_j$, $0\le a_i,b_j<1$,
\begin{equation}\label{4.1}
\mathbb{P}[w(i,j)=m]=(1-a_ib_j)(a_ib_j)^m,
\end{equation}
$m\ge 0$. Also, we set $w(0,j)=w(j,0)=0$, $j\ge 0$. Define as previously, (\ref{2.9}),
\begin{equation}\label{4.2}
G(i,j)=\max (G(i-1,j),G(i,j-1))+w(i,j),
\end{equation}
$(i,j)\in\mathbb{Z}_+^2$, where $G(i,0)=G(0,i)=0$, $i\ge 0$. 

Given a partition, or 
right/down path $\lambda$, set $W(\lambda)=(w(i,j))_{(i,j)\in S(\lambda)}$, where
$S(\lambda)$ is the shape of $\lambda$ as defined previously. If $\lambda=
(M,\dots,M,0,\dots)$ with $\ell(\lambda)=N$, then $W(\lambda)$ is the $M\times N$-matrix
$(w(i,j))_{1\le i\le M,1\le j\le N}$. We want to map $W(\lambda)$ to a family of 
non-intersecting paths in a weighted graph $\mathcal{G}(\lambda)$, in such a way that the 
top path gives the values $(G(P_j(\lambda)))_{j=0}^{K+L}$ of $G(i,j)$ along the
right/down path.

Corresponding to (\ref{4.1}) we define the {\it weight} of $W(\lambda)$ by
\begin{equation}\label{4.3}
\prod_{(i,j)\in S(\lambda)}(a_ib_j)^{w(i,j)}.
\end{equation}
We also want the mapping to the non-intersecting paths to be weight-preserving so that 
we can use the paths to study $G(M,N)$.

The directed graph $\mathcal{G}(\lambda)$ is defined as follows. The vertices are
$\{-L,-L+1,\dots,K\}\times \mathbb{Z}$, and the undirected edges connect $(i,j)$, $(i+1,j)$
for $i=-L,\dots, K-1$, $j\in\mathbb{Z}$ (horizontal edges) and
$(i,j)$, $(i, j+1)$ for $i=-L+1,\dots, K$, $j\in\mathbb{Z}$ (vertical edges).
The step $P_j(\lambda)P_{j+1}(\lambda)$ in the right/down path $\lambda$ is a
{\it right-step} if $P_j(\lambda)=(i-1,x)$ and $P_{j+1}(\lambda)=(i,x)$. In that case the 
vertical edges with first coordinate $-L+j+1$ are directed from $(-L+j+1,k)$ to
$(-L+j+1,k+1)$, $k\in\mathbb{Z}$, i.e upwards, and are given the weight $a_i$ 
({\it up-edges}). The step $P_j(\lambda)P_{j+1}(\lambda)$ is a {\it down-step} if
$P_j(\lambda)=(x,i)$ and $P_{j+1}(\lambda)=(x,i-1)$. In that case the vertical edges 
with first coordinate $-L+j+1$ are directed from $(-L+j+1,k)$ to
$(-L+j+1,k-1)$, $k\in\mathbb{Z}$, i.e downwards, and are given the weight $b_i$ 
({\it down-edges}). All horizontal edges are directed to the right.

To a path $\pi$ in $\mathcal{G}(\lambda)$ from $(-L,1-j)$ to $(K,1-j)$, for some $j\ge 1$, 
we can associate points $Q_i(\pi)=(i-L,x_i)$, $0\le i\le K+L$. We let
$Q_i(\pi)$ be the last vertex in the directed path with first coordinate $-L+i$.
If the edges on the vertical line $x=-L+i$ 
are up-edges then $x_i\ge x_{i-1}$, if they are down-edges,
then $x_i\le x_{i-1}$. We have $x_0=x_{L+L}=1-j$.

We can now formulate a theorem which gives the mapping from $W(\lambda)$ to non-intersecting
paths in $\mathcal{G}(\lambda)$. To our knowledge this theorem has not appeared in 
its present form in the literature so we will give a proof in section \ref{Corner}.
For an investigation of the behaviour along right/down paths in the Poissonian case
see \cite{BoOl}.

\begin{theorem}\label{Thm4.1}
Let $\lambda$ be a partition giving a right/down path $(P_j(\lambda))_{j=0}^{K+L}$,
$L=\ell(\lambda)$, $K=\lambda_1$, from $(0,L)$ to $(K,0)$. There is a 
one-to-one weight preserving mapping from $W(\lambda)$ with weight (\ref{4.3}) to
non-intersecting paths $(\pi_1,\pi_2,\dots)$ in the weighted directed graph
$\mathcal{G}(\lambda)$, where $\pi_j$ goes from $(-L,1-j)$ to $(K,1-j)$,
$j\ge 1$. The path $\pi_j$ consists only of horizontal edges if $j\ge \min(K,L)$.
If $Q_i(\pi_1)=(i-L,x_i)$, $0\le i\le K+L$, are the points associated to the top
path $\pi_1$, then $x_i=G(P_i(\lambda))$.
\end{theorem}

To the paths $\pi_1,\pi_2,\dots$ we can associate a point configuration
$(r,x^r_j)$, $-L\le r\le K$, $j\ge 1$, by letting $Q_i(\pi_j)=(i=L,x^{i-L}_j)$. Note that
$x^{-L}_j=x^K_j=1-j$, $j\ge 1$ are fixed. Also, $x^r_j=1-j$, $-L\le r\le K$, if $j>
\min(K,L)$. Hence theorem \ref{Thm4.1} maps $W(\lambda)$ with probability measure
(\ref{4.1}) to a point process in $\{-L+1,\dots,K-1\}\times\mathbb{Z}$.
By the general formalism presented above this will be a determinantal point process.

\begin{theorem}\label{Thm4.2}
Let $\lambda$ be a partition, a right/down path from $(L,0)$ to $(0,K)$, $L=\ell(\lambda)$,
$K=\lambda_1$. The probability measure (\ref{4.1}) on $W(\lambda)$ can be mapped to a 
determinantal point process on $\{-L+1,\dots,K-1\}\times\mathbb{Z}$. Set
\begin{equation}\label{4.4}
f^+_r(z)=\frac{1-a_i}{1-a_iz}\quad,\quad f^-_r(z)=1
\end{equation}
if the edges on $x=r+1$ are up-edges with weight $a_i$, and
\begin{equation}\label{4.5}
f^-_r(z)=1\quad,\quad f^-_r(z)=\frac{1-b_i}{1-b_i/z}
\end{equation}
if the edges on $x=r+1$ are down-edges with weight $b_i$, $-L\le r<K$. The
correlation kernel is then given by
\begin{equation}\label{4.6}
K_\lambda(r,x;s,y)=-\phi_{r,s}(x,y)+
\frac 1{(2\pi i)^2}\int_{\gamma_{r_2}}\frac {dz}z\int_{\gamma_{r_1}}\frac {dw}w
\frac{w^y}{z^x}\frac{z}{z-w} F_\lambda(z,w),
\end{equation}
where $\max(b_i)<r_1<r_2<\min(1/a_i)$,
\begin{equation}\label{4.7}
F_\lambda(z,w)=\prod_{t=r}^{K-1}f^-_t(\frac 1z)\prod_{t=-L}^{s-1}f^+_t(\frac 1w)
\prod_{t=-L}^{r-1}f^+_t(\frac 1z)^{-1}\prod_{t=s}^{K-1}f^-_t(\frac 1w)^{-1},
\end{equation}
$\phi_{r,s}\equiv 0$ if $r\ge s$ and
\begin{equation}\label{4.8}
\phi_{r,s}(x,y)=\frac 1{2\pi}\int_{-\pi}^\pi e^{i(y-x)}F_\lambda(e^{i\theta},e^{i\theta})
d\theta
\end{equation}
if $r<s$.
\end{theorem}

\begin{proof}
(Sketch) The transition weight to go from $(r,x)$ to $(r+1,y)$ in $\mathcal{G}(\lambda)$ is
\begin{equation}\label{4.9}
\phi_{r,r+1}(x,y)=
\begin{cases} a_i^{y-x} &\text{if $y\ge x$}\\
0  &\text{if $y<x$}\\
\end{cases}
\end{equation}
if the vertical edges on the line $r+1$ are up-edges with weight
$a_i$, and
\begin{equation}\label{4.10}
\phi_{r,r+1}(x,y)=
\begin{cases} b_i^{x-y} &\text{if $\ge y$}\\
0  &\text{if $x<y$}\\
\end{cases}
\end{equation}
if they are down-edges with weight $b_i$. It follows from theorem \ref{Thm4.1} and
the LGV-theorem that the probability measure on the point configuration is given by 
(\ref{1.41}). Hence it has determinantal correlation functions by proposition \ref{Prop1.9}.
We have $\phi_{r,r+1}(x,y)=\hat{f}_r(y-x)$, where $f_r=f^+_rf^-_r$. The matrix $A$ in (\ref
{1.45}) is a Toeplitz matrix and we can use the Wiener-Hopf factorization technique in 
section \ref{Asymptotics} to see that the correlation kernel is given by (\ref{4.6}),
compare (\ref{3.8}).
\end{proof}

\subsection{The Schur and Plancherel measures}\label{Subsect4.2}

Take $\lambda=(N,\dots, N,0,\dots)$, $\ell(\lambda)=N$. Then we have only up-edges to the 
left of the origin and down-edges to the right of the origin. Furthermore
$G(N,N)=\max_{j\ge 1} x_j^0$. Restrict the attention to the point process $x^0_j$,
$j\ge 1$, above the origin. This is then a determinantal point process with 
correlation kernel
\begin{equation}\label{4.9'}
K_N(x,y)=\frac 1{(2\pi i)^2}\int_{\gamma_{r_2}}\frac {dz}z\int_{\gamma_{r_1}}\frac {dw}w
\frac{w^y}{z^x}\frac{z}{z-w}\prod_{j=1}^N\frac{(1-a_j/z)(1-b_jw)}{(1-a_j/w)(1-b_jz)}.
\end{equation}
Consider now the case $a_i=\sqrt{q}$, $1\le i\le N$, $b_i=\sqrt{q}$, $1\le i\le M$,
$b_i=0$, $M<i\le N$, where $M\le N$. The kernel can then be expressed in terms of
Meixner polynomials, \cite{JoSh}, and is called the {\it Meixner kernel}. If we scale
appropriately around the last particle this kernel has the Airy kernel as its scaling
limit and proceeding as we did for the Krawtchouk ensemble we can prove the following
theorem, \cite{JoSh}

\begin{theorem}\label{Thm4.3}
Let $0<q<1$ and consider $G(M,N)$ defined by (\ref{4.2}), where $w(i,j)$ are i.i.d.
geometric random variables with parameter $q$. Then, for $\gamma\ge 1$,
\begin{equation}
\lim_{N\to\infty}\mathbb{P}[\frac{G([\gamma N],N)-N\omega(\gamma,q)}{N^{1/3}\sigma(
\gamma,q)}\le s]=F_{\text{TW}}(s)
\notag
\end{equation}
for any $s\in\mathbb{R}$, where
\begin{equation}
\omega(\gamma,q)=\frac{(1+\sqrt{q\gamma})^2}{1-q}-1,
\notag
\end{equation}
\begin{equation}
\sigma(\gamma,q)=\frac{q^{-1/6}\gamma^{-1/6}}{1-q}(\sqrt{\gamma}+\sqrt{q})^{2/3}
(1+\sqrt{q\gamma})^{2/3}.
\notag
\end{equation}
\end{theorem}

The probability measure on the points $x^0_1>\dots>x^0_N$ ($M=N$) is the so-called {\it 
Schur measure}, \cite{Ok}. The points $x^0_1>\dots>x^0_N$ can be related to a partition 
$\mu$ by $x^0_i=\mu_i-i+1$ so we can also think about the Schur measure as a measure on 
partitions. This is the probability measure on partitions $\mu$ that we obtain if we map
$(w(i,j))$, an $N\times N$ matrix, $w(i,j)$ as in (\ref{4.1}), 
to a pair of semi-standard Young tableaux with shape $\mu$ using the RSK-correspondence, 
\cite{Sa}. We will not prove this here. In the present case all the vertical edges 
to the left of the origin and at the origin in the
directed graph are up-edges, whereas those to the right of the origin are down-edges.
The non-intersecting paths pass through the points $\mu_i-i+1$, $1\le i\le N$,
above the origin. By the LGV-theorem the Schur measure can thus be written
\begin{equation}\label{4.12}
p_S(\mu)=\frac 1{Z_N}\det(\phi(1-j,\mu_i-i+1))\det(\psi(1-j,\mu_i-i+1)),
\end{equation}
where
\begin{align}
\phi(u,v)&=\sum_{1\le i_1<\dots<i_{v-u}\le N}a_{i_1}\dots a_{i_{v-u}}=
h_{v-u}(a_1,\dots,a_N),
\notag\\
\psi(u,v)&=\sum_{1\le i_1<\dots<i_{v-u}\le N}b_{i_1}\dots b_{i_{v-u}}=
h_{v-u}(b_1,\dots,b_N)
\notag
\end{align}
Here $h_k(a_1,\dots,a_N)$ is the $k$'th complete symmetric polynomial. The symmetric
polynomial
\begin{equation}\label{4.13}
s_\mu(a_1,\dots,a_N)=\det(h_{\mu_i-i+j}(a_1,\dots,a_N))
\end{equation}
is the {\it Schur polynomial} labelled by $\mu$. The Schur measure can thus be written
\begin{equation}\label{4.14}
p_S(\mu)=\frac 1{Z_N}s_\mu(a_1,\dots, a_N)s_\mu(b_1,\dots,b_N),
\end{equation}
which explains the name. The normalization is, \cite{Sa},
\begin{equation}\label{4.15}
Z_N=\sum_\mu s_\mu(a_1,\dots, a_N)s_\mu(b_1,\dots,b_N)=\prod_{i,j=1}^N\frac 1{1-a_ib_j}.
\end{equation}
It follows from above that, under the Schur measure, $\sum_{i\ge 1}\delta_{\lambda_i-i}$
is a determinantal point process on $\mathbb{Z}$ with correlation kernel (\ref{4.9'}).

If we restrict our attention to the case when $W=(w(i,j))_{1\le i,j\le N}$ is a permutation
matrix, then $G(N,N)=\ell_N(\sigma)$ is exactly the length of the longest increasing
subsequence in the permutation corresponding to $W$. Restricting to a permutation
matrix means that we want to have exactly one up-step of size 1 on each line with
up-edges and exactly one down-step of size 1 on each line with down-edges. If the height 
configuration at the origin is $\mu_i-i+1$, $i\ge 1$, $\ell_N(\sigma)=\mu_1$, the
corresponding measure on partitions $\mu$ is
\begin{equation}\label{4.16}
\mathbb{P}_{\text{Plan\,},N}[\mu]=\frac 1{N!}[a_1\dots a_Nb_1\dots b_N]s_\mu(a)s_\mu(b),
\end{equation}
where $[a_1\dots a_Nb_1\dots b_N]$ means that we take the coefficient of the monomial 
\newline
$a_1\dots a_Nb_1\dots b_N$ in $s_\mu(a)s_\mu(b)$. This measure is called the
{\it Plancherel measure}, and is also given by $f_\mu^2/N!$, where $f_\mu$ is the
number of standard Young tableaux with shape $\mu$. We saw above that when studying the
problem of the length of the longest increasing subsequence in a uniform random
permutation from $S_N$ it was natural to let $N$ be a Poisson($\alpha$) random
variable. We thus consider the {\it Poissonized Plancherel measure},
\begin{equation}\label{4.17}
\mathbb{P}_{\text{PP}}^\alpha [\mu]=\sum_{N=0}^\infty\frac{\alpha^N}{N!}
\mathbb{P}_{\text{Plan\,},N}[\mu],
\end{equation}
where $\mathbb{P}_{\text{Plan\,},N}[\mu]=0$ if $N$ is not a partition of $N$.

\begin{theorem}\label{Thm4.4} (\cite{BOO},\cite{JDOPE}). Under the map
$\mu\to\sum_{i\ge 1}\delta_{\mu_i-i}$, the Poissonized Plancherel measure is mapped
to a determinantal point process with correlation kernel $B^\alpha(x,y)$, the 
discrete Bessel kernel, given by (\ref{3.34}).
\end{theorem}
\begin{proof} 
(Sketch). Suppose that $g:\mathbb{Z}\to\mathbb{C}$ has support in $[-L,\infty)\cup
\mathbb{Z}$ for some $L\ge 0$. By (\ref{4.14}) - (\ref{4.17}),
\begin{align}\label{4.18}
&\sum_\mu \prod_{j=1}^\infty (1+g(\mu_j-j))\mathbb{P}_{\text{PP}}^\alpha [\mu]
\notag\\
&=\sum_{N=0}^\infty\frac{\alpha^N}{(N!)^2}[a_1\dots a_Nb_1\dots b_N]
\prod_{i,j=1}^N\frac 1{1-a_ib_j}\sum_\mu \prod_{j=1}^\infty (1+g(\mu_j-j))p_S(\mu).
\end{align}
Now, since the Schur measure has determinantal correlation functions we know that
\begin{equation}\label{4.19}
\sum_\mu \prod_{j=1}^\infty (1+g(\mu_j-j))p_S(\mu)
=\sum_{k=0}^\infty\frac 1{k!}\sum_{x\in\mathbb{Z}^k}\prod_{j=1}^k g(x_j)
\det(K_N(x_i+1,x_j+1))_{1\le i,j\le k},
\end{equation}
where $K_N$ is given by (\ref{4.9'}). Inserting (\ref{4.19}) into (\ref{4.18}) a
rather long computation, which we omit, gives
\begin{equation}
\sum_\mu \prod_{j=1}^\infty (1+g(\mu_j-j))\mathbb{P}_{\text{PP}}^\alpha [\mu]
=\det(I+B^\alpha g)_{\ell^2(\mathbb{Z})},
\notag
\end{equation}
which proves the theorem by proposition \ref{Prop1.4}.
\end{proof}

\subsection{A discrete polynuclear growth model}\label{Subsect4.3}

Consider now the right/down path which is given by $\lambda=(N,N-1,\dots,1,0,\dots)$
and choose $a_i=b_i=\sqrt{q}$. set $G(i+1/2,j+1/2)=G(i,j)$, $i,j\ge 0$, and
\begin{equation}\label{4.20}
h(x,t)=G(\frac{t+x+1}2,\frac{t-x+1}2)
\end{equation}
for $x\in\mathbb{Z}$, $t\ge 0$, $|x|\le t$ and $h(x,t)=0$ if $|x|>t$. We see that 
$h(x,N)$, $x=-N,\dots,N$ are exactly the values of $G(i,j)$ along the right down
path $\lambda$. Set $\omega(x,t)=0$ if $t-x$ is even or if $|x|>t$, and
\begin{equation}\label{4.21}
\omega(x,t)=w(\frac{t+x+1}2,\frac{t-x+1}2)
\end{equation}
otherwise. From (\ref{4.2}) it follows that
\begin{equation}\label{4.22}
h(x,t+1)=\max(h(x-1,t), h(x,t),h(x+1,t))+\omega(x,t+1).
\end{equation}
We think of $h(x,t)$ as the height above $x$ at time $t$. This growth model is a
{\it discrete polynuclear growth model}, \cite{KrSp}. It is different from the
one related to the shuffling procedure in the Aztec diamond. It follows from theorem 
\ref{Thm4.2} that the height fluctuations above $[ct]$, $0\le c<1$ at time $t$ are of
order $t^{1/3}$ for large $t$ and are described by the Tracy-Widom distribution.
But we actually know the whole extended kernel (\ref{4.6}) for
$\lambda=(N,N-1,\dots)$ and hence we can study $x\to h(x,t)$, $|x|\le t$, as a process.
This was done in \cite{PrSp} for the Poissonian limit of the model corresponding to
random permutations, where we have an extended discrete Bessel kernel.

If we consider the process at time $2N-1$, and look at even $x$, we are studying the process 
$u\to G(N+u,N-u)$, $|u|<N$. The appropriate scaling limit of the kernel (\ref{4.6}) will
again be the extended Airy kernel, similarly to what we got when we studied
the NPR-boundary process. We have the following theorem, \cite{JDPNG}, which we will not
prove. Basically we have again to investigate the convergence of the extended kernel
(\ref{4.6}) to the extended Airy kernel using the saddle-point method. We define the rescaled
process $H_N(t)$ by
\begin{equation}\label{4.23}
G(N+u,N-u)=\frac{2\sqrt{q}}{1-\sqrt{q}}N+dN^{1/3}H_N\left(\frac{1-\sqrt{q}}{1+\sqrt{q}}
dN^{-2/3}u\right),
\end{equation}
where
\begin{equation}\label{4.24}
d=\frac{(\sqrt{q})^{1/3}(1+\sqrt{q})^{1/3}}{1-q}.
\end{equation}
We extend it to a continuous process defined for all times by linear interpolation.

\begin{theorem}\label{Thm4.5}
We have $H_N(t)\to\mathcal{A}(t)-t^2$, where $\mathcal{A}(t)$ is the Airy process,
in the sense of convergence of finite-dimensional distributions.
\end{theorem}

From (\ref{2.12}) we know that $G(N,N)$ is a certain point-to-point last passage
time. It is also natural to consider a {\it point-to-line last passage time},
\begin{equation}\label{4.25}
G_{\text{pl}}(N)=\max_{|u|<N} G(N+u,N-u),
\end{equation}
i.e. we take the maximum of $\sum_{(i,j)\in\pi}$ over all up/right paths from
$(1,1)$ to the line $x+y=2N$. A maximal path will be like a directed polymer with one free 
end. From (\ref{4.23}) we see that it is natural to consider the maximum of the
process $H_N(t)$. To do so we need a stronger form of convergence then in theorem 
\ref{Thm4.5}. The next theorem is proved in \cite{JDPNG}.

\begin{theorem}\label{Thm4.6}
There is a continuous version of the Airy process $\mathcal{A}(t)$ and $H_N(t)\to
\mathcal{A}(t)-t^2$ as $N\to\infty$ in the weak star topology of probability
measures on $C(-T,T)$ for any fixed $T$.
\end{theorem}

It is proved in \cite{BaRa} that 
\begin{equation}\label{4.26}
\mathbb{P}[\frac{G_{\text{pl}}(N)-2\sqrt{q}(1-\sqrt{q})^{-1}N}{dN^{1/3}}\le s]
\to F_1(s)
\end{equation}
as $N\to \infty$, where $F_1$ is the largest eigenvalue, Tracy-Widom law, for the
Gaussian Orthogonal Ensemble (GOE). This law is different from $F_{TW}$, which is
often denoted by $F_2$, \cite{TrWi}. We will not give its explicit form here. If we 
combine this result with theorem \ref{Thm4.6} we see that
\begin{equation}\label{4.27}
F_1(s)=\mathbb{P}[\sup_{t\in\mathbb{R}}(\mathcal{A}(t)-t^2)\le s].
\end{equation}
It would be interesting to have a more direct approach to this result.

The maximal path in the point-to-line last passage problem is not necessarily
unique so there could be several possible endpoints on the line. Set
\begin{equation}\label{4.28}
K_N=\inf\{ s\,;\,\sup_{t\le s}H_N(t)=\sup_{t\in\mathbb{R}} H_N(t)\},
\end{equation}
and similarly for the limiting process
\begin{equation}\label{4.29}
K=\inf\{ s\,;\,\sup_{t\le s}(\mathcal{A}(t)-t^2)=\sup_{t\in\mathbb{R}} 
 (\mathcal{A}(t)-t^2)\}.
\end{equation}
If we could show that the process $t\to\mathcal{A}(t)-t^2$ has a unique point of maximum
almost surely it would follow that $K_N\to K$, and the law of $K$ would be the law
of transversal fluctuations of the endpoint of a maximal path in a point-to-line
problem. However, the above argument gives us no clue what this law could be. 
For all we know it could be Gaussian.

\subsection{Proof of theorem \ref{4.1}}\label{Subsect4.4}

The procedure is very close to the Robinson-Schensted-Knuth correspondence. See
\cite{OCon} for a related analysis. 

We have to define the paths $\pi_1,\pi_2,\dots$\,. Set $w^{(0)}(i,j)=w(i,j)$. 
Assume that we have defined $w^{(k)}(i,j)$ for some $k\ge 0$, $i,j\ge 0$. Set
$G^{(k)}(0,j)=G^{(k)}(j,0)=0$, $j\ge 0$, and
\begin{equation}\label{4.30}
G^{(k)}(i,j)=\max(G^{(k)}(i-1,j),G^{(k)}(i,j-1))+w^{(k)}(i,j),
\end{equation}
$i,j\ge 1$, so that $G^{(0)}(i,j)=G(i,j)$. Also, define
\begin{equation}\label{4.31}
w^{(k+1)}(i,j)=\min(G^{(k)}(i-1,j),G^{(k)}(i,j-1))-G^{(k)}(i-1,j-1).
\end{equation}
This defines $w^{(k)}(i,j)$ and $G^{(k)}(i,j)$ recursively for all $k\ge 0$,
$i,j\ge 1$.

The path $\pi_k$ goes between the points $(-L+j,G^{(k-1)}(P_j(\lambda))-k+1)$,
$j=0,\dots,K+L$ and respects the direction of the edges in $\mathcal{G}(\lambda)$.

\begin{claim}\label{Cl4.7}
We have that $w^{(k)}(i,j)=0$ if $0\le i\le k$ or $0\le j\le k$.
\end{claim}

To prove the claim we use induction on $k$. The claim is true by definition if
$k=0$. Now, if $w^{(k)}(i,j)=0$ for $0\le i\le k$ or $0\le j\le k$, then 
$G^{(k)}(i,j)=0$ for $0\le i\le k$ or $0\le j\le k$. If $k\le k+1$ or $j\le k+1$, then
$i-1\le k$ or $j-1\le k$. Hence, $G^{(k)}(i-1,j)=0$ or $G^{(k)}(i,j-1)=0$, so
$w^{(k+1)}(i,j)=0$ by (\ref{4.31}). This proves claim \ref{Cl4.7}.

\begin{claim}\label{Cl4.8}
The paths $\pi_1,\pi_2,\dots$ in $\mathcal{G}(\lambda)$ do not intersect.
\end{claim}

We first prove that 
\begin{equation}\label{4.32}
\min(G^{(k)}(i-1,j),G^{(k)}(i,j-1))\ge G^{(k+1)}(i,j)
\end{equation}
for all $i,j\ge 1$, $k\ge 0$. If $(i,j)=(1,1)$, then (\ref{4.32}) is clearly true since
$G^{(k+1)}(1,1)=0$ for all $k\ge 0$ by claim \ref{Cl4.7}. Assume (A) that (\ref{4.32})
holds for $1\le i,j\le m$. If $i=m+1$ and $j=1$, then (\ref{4.32}) holds because
$G^{(k+1)}(m+1,1)=0$ by claim \ref{Cl4.7}. Assume (B) that (\ref{4.32}) holds for
$i=m+2$, $1\le j<n$, where $n\le m$. We want to prove (\ref{4.32}) for $i=m+1$, $j=n$. By
(\ref{4.30}) and (\ref{4.31}) the inequality (\ref{4.32}) with $i=m+1$, $j=n$ is equivalent
to
\begin{align}
&\min(G^{(k)}(m+1,n-1),G^{(k)}(m,n))\ge\max(G^{(k+1)}(m,n),G^{(k+1)}(m+1,n-1))
\notag\\
&+\min(G^{(k)}(m+1,n-1),G^{(k)}(m,n))
-G^{(k)}(m,n-1)
\notag
\end{align}
i.e.
\begin{equation}\label{4.33}
G^{(k)}(m,n-1)\ge\max(G^{(k+1)}(m,n),G^{(k+1)}(m+1,n-1)).
\end{equation}
By assumption (A), $G^{(k)}(m,n-1)\ge G^{(k+1)}(m,n)$ and by assumption (B)
$G^{(k)}(m,n-1)\ge G^{(k+1)}(m+1,n-1)$, so (\ref{4.33}) holds.

By induction (\ref{4.32}) also holds for $i=m+1$, $1\le j\le m$. A similar argument shows 
that  (\ref{4.32}) holds for $1\le i\le m$, $j=m+1$. It remains to consider the case
$(i,j)=(m+1,m+1)$. The (\ref{4.32}) is equivalent to
\begin{equation}\label{4.34}
G^{(k)}(m,m)\ge\max(G^{(k+1)}(m,m+1),G^{(k+1)}(m+1,m))
\end{equation}
by (\ref{4.30}) and (\ref{4.31}). Now $i=m$, $j=m+1$ in (\ref{4.32}) gives
$G^{(k)}(m,m)\ge G^{(k+1)}(m,m+1)$, and $i=m+1$, $j=m$ in (\ref{4.32}) gives
$G^{(k)}(m,m)\ge G^{(k+1)}(m+1,m)$. Hence (\ref{4.34}) holds and we have shown that 
(\ref{4.32}) holds for $1\le i,j\le m+1$ and hence holds for all $i,j\ge 1$ by
induction.

We can use (\ref{4.32}) to prove claim \ref{Cl4.8}. Consider an up-step $P_j(\lambda)$
to $P_{j+1}(\lambda)$, $P_j(\lambda)=(i-1,x)$,  $P_{j+1}(\lambda)=(i,x)$. The path $\pi_k$
then goes from $(-L+j,G^{(k-1)}(P_j(\lambda))-k+1)$ to
$(-L+j+1,G^{(k-1)}(P_{j+1}(\lambda))-k+1)$ via the points $(-L+j+1,m)$, where
$G^{(k-1)}(P_j(\lambda))-k+1\le m\le G^{(k-1)}(P_{j+1}(\lambda))-k+1$. Non-intersection
between $\pi_k$ and $\pi_{k+1}$ holds if
$G^{(k-1)}(P_j(\lambda))-k+1>G^{(k)}(P_{j+1}(\lambda))-k$ or
$G^{(k-1)}(i-1,x)\ge G^{(k)}(i,x)$, which follows from (\ref{4.32}). A down-step
is treated analogously. This proves claim \ref{Cl4.8}.

\begin{claim}\label{Cl4.9}
The path $\pi_k$ is horizontal if $k>\min(K,L)$.
\end{claim}

Assume that $K\le L$ and $k>K$. the path $\pi_k$ goes through the points 
$(-L+j,G^{(k-1)}(P_j(\lambda))-k+1)$. here $P_j(\lambda=(x,y)$ where $x\le K$. Now,
$G^{(k-1)}(x,y)=0$ if $x$ or $y$ is $<k$ by claim \ref{Cl4.7}, so $\pi_k$ goes through
the points $(-L+j,-k+1)$ and is horizontal.

\begin{claim}\label{Cl4.10}
Set $W^{(k)}(\lambda)=(w^{(k)}(i,j))_{(i,j)\in S(\lambda)}$ and define the weight of 
$W^{(k)}(\lambda)$ to be
\begin{equation}
\prod_{(i,j)\in S(\lambda)} (a_ib_j)^{w^{(k)}(i,j)}.
\notag
\end{equation}
Then, the weight of $W^{(k-1)}(\lambda)$ is equal to the weight of $W^{(k)}(\lambda)$ times
the weight of $\pi_k$ in $\mathcal{G}(\lambda)$ for all $k\ge 1$.
\end{claim}

Let $1\le i\le K$. In the weight of $W^{(k-1)}(\lambda)$, $a_i$ enters as
\begin{equation}
a_i^{\sum_{j=1}^m w^{(k-1)}(i,j)}
\notag
\end{equation}
if the up-step with weight $a_i$ is the step from $(i-1,m)$ to $(i,m)$ in the
right/down path $\lambda$. The up-step has size $G^{(k-1)}(i,m)-G^{(k-1)}(i-1,m)$
so we want to show that
\begin{equation}\label{4.35}
G^{(k-1)}(i,m)-G^{(k-1)}(i-1,m)+\sum_{j=1}^m w^{(k)}(i,j)=\sum_{j=1}^m w^{(k-1)}(i,j).
\end{equation}
Now, by (\ref{4.30}) and (\ref{4.31}),
\begin{align}
w^{(k-1)}(i,j)&=G^{(k-1)}(i,j)-\max(G^{(k-1)}(i-1,j),G^{(k-1)}(i,j-1))
\notag\\
-w^{(k)}(i,j)&=G^{(k-1)}(i-1,j-1)-\min(G^{(k-1)}(i-1,j),G^{(k-1)}(i,j-1)).
\notag
\end{align}
Adding these two inequalities and summing over $j$ we obtain
\begin{align}
&\sum_{j=1}^m(w^{(k-1)}(i,j)-w^{(k)}(i,j))=
\sum_{j=1}^m[G^{(k-1)}(i,j)-G^{(k-1)}(i-1,j)]
\notag\\
&-\sum_{j=0}^{m-1}[G^{(k-1)}(i,j)-G^{(k-1)}(i-1,j)]=
G^{(k-1)}(i,m)-G^{(k-1)}(i-1,m-1),
\notag
\end{align}
so we have established (\ref{4.35}). The argument for $b_i$ is analogous and we have proved
claim \ref{Cl4.10}.

To prove the theorem it remains to show that the mapping from $W(\lambda)$ to
$(\pi_1,\pi_2,\dots)$ is one-to-one and weight preserving. That the map is
weight preserving follows by claim \ref{Cl4.9} and repeated use of claim \ref{Cl4.10}.
We have to show that the map is invertible, i.e. given $(\pi_1,\pi_2,\dots)$
we can reconstruct $W(\lambda)$.

If $k>\min(K,L)$, then $W^{(k)}(\lambda)=0$ by claim \ref{Cl4.7}. Assume that we know
$W^{(k)}(\lambda)$ for some $k\ge 1$. We want to show that we can reconstruct
$W^{(k-1)}(\lambda)$. Repeating this we eventually get $W^{(0)}(\lambda)=W(\lambda)$.

Let $\lambda$ be a partition of $N$ and let $\lambda=\lambda^0>\lambda^1>\dots
>\lambda^{N-1}>\lambda^N=\emptyset$ be a sequence of partitions such that we get $\lambda^k$ 
from $\lambda^{k-1}$ be removing one point on the boundary. 
(We have a path from $\lambda$ to $\emptyset$ in the Young lattice, \cite{Sa}.)
Assume that we know all the values of $G^{(k-1)}(m,n)$ for $(m,n)$ along the
boundary of $\lambda^j$. From $\pi_k$ we know all the values of $G^{(k-1)}(m,n)$
along the boundary of $\lambda^0=\lambda$. Let $(m,n)$ be the point 
on the boundary of $\lambda^j$ that we remove when we go to $\lambda^{j+1}$.
Since $\lambda^{j+1}$is also a partition, $(m-1,n)$ and $(m,n-1)$ must also be points
along the boundary of $\lambda^j$. Hence, we know $G^{(k-1)}(m-1,n)$ and
$G^{(k-1)}(m,n-1)$. We can now get $w^{(k-1)}(m,n)$ from (\ref{4.30}). Since we
know $w^{(k)}(m,n)$ we can also get $G^{(k-1)}(m-1,n-1)$ from (\ref{4.31}). hence we know the
values of $G^{(k-1)}(m,n)$ along the boundary of $\lambda^{j+1}$. Proceeding in this way we 
succesively get the numbers $w^{(k-1)}(m,n)$ for $(m,n)\in S(\lambda)$.
This completes the proof of theorem \ref{Thm4.1}.

\medskip\noindent
{\bf Acknowledgement}: I thank the organizers of the Les Houches summer
school on Mathematical Statistical Mechanics for the invitation to present
this series of lectures.


\begin{thebibliography}{99}

\itemsep=\smallskipamount

\bibitem{BDJ} J. Baik, P. A. Deift, K. Johansson, {\em On the distribution
of the length of the longest 
increasing subsequence in a random permutation,} 
J. Amer. Math. Soc., {\bf 12}, (1999), 1119 - 1178

\bibitem{BKMP} J. Baik, T. Kriecherbauer, K.D.T.-R MacLaughlin, P. Miller,
{\em Uniform asymptotics for polynomials orthogonal with respect to 
a general class of discrete weights and universality results
for associated ensembles: announcement of results,}
Int. Math. res. Not., {\bf np. 15} (2003), 821--858

\bibitem{BaRa} J. Baik, E. Rains, {\em Symmetrized random permutations},
in Random Matrix Models and Their Apllications, eds. P.M. Bleher,
A.R. Its, MSRI Publications 40, Cambridge 2001

\bibitem{BGS} O. Bohigas, M. J. Giannoni, C. Schmit, {\em Characterization
of chaotic quantum spectra and universality of level fluctuation laws,}
Phys. Rev. Lett. {\bf 52} (1984) 1-4

\bibitem{BOO} A. Borodin, A. Okounkov \& G. Olshanski, {\em Asymptotics of
Plancherel measures for symmetric groups,} J. Amer. Math. Soc. {\bf
13} (2000), 481--515

\bibitem{BoOl} A. Borodin \& G. Olshanski, {\em Stochastic dynamics related to
Plancherel measure on partitions,} arXiv:math-ph/0402064

\bibitem{CEP} H. Cohn, N. Elkies, J. Propp, {\em Local statistics for 
random domino
tilings of the Aztec diamond,}, Duke Math. J., {\bf 85}, (1996), 117
- 166

\bibitem{CKP} H. Cohn, R. Kenyon, J. Propp, {\em 
A variational principle for domino tilings,} J. Amer. Math. Soc.,
{\bf 14} (2001), 297-346

\bibitem{DV-J} Daley, D.J., Vere-Jones, D., {\em An Introduction to the Theory of
Point Processes,} Vol.1, 2nd ed, Springer (2003)

\bibitem{EKLP} N. Elkies, G. Kuperberg, M. Larsen, J. Propp, {\em
Alternating-Sign Matrices and Domino Tilings (Part I)} 
and {\em Alternating-Sign Matrices and Domino Tilings (Part II),} 
J. of Algebraic Combin., {\bf 1}, (1992), 111- 132 and 219 - 234

\bibitem{EyMe} B. Eynard, M.L. Mehta, {\em Matrices coupled in a chain
    I: Eigenvalue correlations}, J. of Phys. A, {\bf 31} (1998), 4449
    - 4456

\bibitem{FS} P. L. Ferrari, H. Spohn, {\em Step fluctuations for a
faceted crystal}, J. Stat. Phys., {\bf 113} (2003), 1 - 46

\bibitem{FNH} P.J. Forrester, T. Nagao, G. Honner, {\em Correlations
    for the orthogonal-unitary and symplectic-unitary transitions at
    the soft and hard edges}, Nucl. Phys. B, {\bf 553} (1999), 601 - 643

\bibitem{GGK} Gohberg, I, Goldberg, S., Krupnik, N., {\em Traces and
Determinants of linear Operators,} Birkh\"auser, Basel (2000)

\bibitem{Gu} Guhr, T., M\"uller-Groeling, A, Weidenm\"uller, H-A, 
{\em Random-matrix theories in quantum physics: common concepts},  
Phys. Rep.  {\bf 299}  (1998),  no. 4-6, 189 - 425. 

\bibitem{JPS} W. Jockush, J. Propp, P. Shor, {\em Random domino tilings
and the arctic circle theorem,} preprint 1995, math.CO/9801068

\bibitem{Jo1} K. Johansson, {\em The longest increasing subsequence in a
random permutation and a unitary random matrix model,}
Math. Res. Lett.,
{\bf 5} (1998), 63 -- 82

\bibitem{JoSh} K. Johansson, {\em Shape fluctuations and random matrices,}
Commun. Math. Phys., {\bf 209}, (2000), 437 - 476

\bibitem{JDOPE} K. Johansson, {\em Discrete orthogonal polynomial ensembles
and the Plancherel measure,} 
Annals of Math., {\bf 153} (2001), 259 - 296

\bibitem{JNIP} K. Johansson, {\em Non-intersecting paths, random
    tilings and random matrices}, 
    Probab.Theory Relat. Fields, {\bf 123} (2002), 225--280

\bibitem{JDPNG} K. Johansson, {\em Discrete polynuclear growth and
determinantal processes,} Commun. Math. Phys., {\bf 242} (2003), 277 - 329

\bibitem{JoAz} K. Johansson, {\em The Arctic circle boundary and the Airy
process,} Ann. of Probab., {\bf 33} (2005), 1 - 30

\bibitem{KaSa} N. M. Katz, P. Sarnak, {\em Zeroes of Zeta Functions
and Symmetry,} Bull. AMS, {\bf 36} (1999), 1 - 26

\bibitem{Ke} R. Kenyon, {\em Local statistics of lattice dimers,}
  Ann. Inst. H. Poincar\'e, Probabilit\'es et Statistiques, {\bf 33}
  (1997), 591 - 618

\bibitem{KrSp} J. Krug, H. Spohn, {\em Kinetic Roughening of Growing
Interfaces}, in Solids far from Equilibrium: Growth, Morphology and
Defects , Ed. C. Godr\`eche, 479 - 582, Cambridge University Press,
1992

\bibitem{Me} M. L. Mehta, {\em Random Matrices,} 2nd ed., Academic Press, 
San Diego 1991

\bibitem{OCon} N. O'Connell, {\em A path-transformation for random walks and the 
Robinson-Schensted correspondence,}  Trans. Amer. Math. Soc. {\bf 355} (2003), 3669 - 3697 

\bibitem{Ok} A. Okounkov, {\it Infinite wedge and random
    partitions,} Selecta Math. (N.S.), {\bf 7} (2001), 57--81

\bibitem{OR} A. Okounkov, N. Reshetikhin, 
{\em Correlation function of Schur process with applications to local
  geometry with application to local geometry of a random
  3-dimensional Young diagram}, J. Amer. Math. Soc., {\bf 16} (2003), 
581 - 603

\bibitem{PrSp} M. Pr\"ahofer, H. Spohn, {\em Scale invariance of the
    PNG droplet and the Airy process}, J. Stat. Phys., {\bf 108}
(2002), 1076--1106

\bibitem{Ro} H. Rost, {\em Non-Equilibrium Behaviour of a Many
    Particle Process: Density Profile and Local Equilibria,}
    Zeitschrift f. Wahrsch. Verw. Geb., {\bf 58} (1981), 41 - 53

\bibitem{Sa} B. Sagan, {\em The Symmetric Group,} 2nd edition, Springer, New York 2001

\bibitem{So} A. Soshnikov, {\em Determinantal random point fields,}
 Russian Math. Surv., {\bf 55} (2000), 923 - 975

\bibitem{St} R. P. Stanley, {\em Enumerative Combinatorics,} Vol. 2,
Cambridge University Press, 1999

\bibitem{Stem} J. R. Stembridge, {\em Nonintersecting Paths,
    Pfaffians, and Plane Partitions,} Adv. in Math., {\bf 83} (1990), 96 - 131

\bibitem{TW1} C. A. Tracy, H. Widom, {\em Level Spacing Distributions and
the Airy Kernel,} Commun. Math. Phys., {\bf 159}, (1994), 151 - 174 

\bibitem{TrWi} C. A. Tracy, H. Widom, {\em On orthogonal and symplectic matrix ensembles,}
Comm. Math. Phys.  {\bf 177} (1996), 727 - 754

\end{thebibliography}
\end{document}